\documentclass[sn-standardnature]{sn-jnl}% Standard Nature Portfolio Reference Style
\usepackage{graphicx}% Include figure files
\usepackage{dcolumn}% Align table columns on decimal point
\usepackage{bm}% bold math
\usepackage{amssymb}
\usepackage{amsmath}
\usepackage{psfrag}
\usepackage{mathtools}
\usepackage{color}
\usepackage{xcolor}
\usepackage[export]{adjustbox}
\usepackage[english]{babel}
\usepackage[T1]{fontenc}
\usepackage{stmaryrd}
\usepackage{graphicx}% Include figure files
\usepackage{dcolumn}% Align table columns on decimal point
\usepackage{bm}% bold math
\usepackage{esint}
\usepackage{framed,xcolor}

\usepackage[utf8]{inputenc}
\usepackage[T1]{fontenc}
\usepackage{mathptmx}
\usepackage{etoolbox}

\definecolor{cmap1}{rgb}{1,0.6,0.6}
\definecolor{cmap2}{rgb}{0.6,0.8,1}
\definecolor{IntenseRed}{rgb}{0.8,0.4,0.4}
\definecolor{IntenseBlue}{rgb}{0.4,0.56,0.8}
\definecolor{vortexyellow}{cmyk}{0.02,0.5,0.47,0}
\definecolor{vortexblue}{cmyk}{0.58,0.56,0.02,0}
\definecolor{myPurple}{RGB}{153,51,204}  
\definecolor{myGreen}{RGB}{51,179,51}
\definecolor{blue1}{cmyk}{0.88,0.77,0,0}
\definecolor{orangex}{cmyk}{0, 0.37,0.97,0}
\definecolor{orangex1}{cmyk}{0, 0.13,0.39,0}
\definecolor{purple1}{cmyk}{0.17,0.53,0,0}
\definecolor{green1}{cmyk}{0.5,0,1,0}
\definecolor{blue2}{cmyk}{1,1,0,0}
\definecolor{orange2}{cmyk}{0,0.5,1,0}
\definecolor{red2}{cmyk}{0,1,1,0}
\definecolor{purple2}{cmyk}{0.35,1,0.35,0.1}
\definecolor{redweak}{cmyk}{0,0.31,0.15,0}
\definecolor{blueweak}{cmyk}{0.23,0.21,0.01,0}
\definecolor{blue2}{cmyk}{0.99,0.94,0.05,0.01}

\definecolor{f1green}{cmyk}{0.76,0.03,0.56,0}
\definecolor{f1red2}{cmyk}{0.01,0.92,0.95,0}
\definecolor{f1red}{cmyk}{0.89,0.29,0.99,0.19}
\definecolor{f1purp2}{cmyk}{0.77,0.99,0.03,0}
\definecolor{f1purp}{cmyk}{0.65,0.58,0.57,0.37}

\definecolor{f1green2}{cmyk}{0.89,0.29,0.99,0.19}
\definecolor{f3blue}{cmyk}{0.88,0.77,0,0}

\definecolor{f2blue}{cmyk}{0.3,0,0.08,0}

\definecolor{SIblue}{cmyk}{0.7,0.15,0,0}
\definecolor{SIorange}{cmyk}{0,0.8,0.92,0}
\definecolor{SIbrown}{cmyk}{0.23,0.39,0.63,0.01}
\definecolor{SIpurple}{cmyk}{0.59,0.9,0,0}
\definecolor{orangex1}{cmyk}{0, 0.13,0.39,0}
\definecolor{blueweak}{cmyk}{0.23,0.21,0.01,0}
\definecolor{purple2}{cmyk}{0.35,1,0.35,0.1}
\definecolor{redweak}{cmyk}{0,0.31,0.15,0}
\definecolor{armorange}{cmyk}{0,0.44,1,0}
\definecolor{swirlpurp}{cmyk}{0.6,0.9,0,0}

\definecolor{Vblue}{cmyk}{0.85,0.5,0,0}
\definecolor{Ured}{cmyk}{0,1,1,0}
\definecolor{iblue}{cmyk}{0.8,0.58,0,0}

\jyear{2026}%

\theoremstyle{thmstyleone}%
\theoremstyle{thmstyletwo}%

\theoremstyle{thmstylethree}%

\begin{document}

\title[Objective detection of coherent vortices from instantaneous flow data]{Objective detection of coherent vortices from instantaneous flow data}

\author*[1]{\fnm{Tiemo} \sur{Pedergnana}}\email{tiemop@mit.edu}

\author[2]{\fnm{Florian} \sur{Kogelbauer}}

\affil[1]{\orgdiv{Department of Mechanical Engineering}, \orgname{Massachusetts Institute of Technology}, \state{MA}, \country{USA}}

\affil[2]{\orgdiv{Department of Mathematics}, \orgname{ETH Zürich}, \state{Zürich}, \country{Switzerland}}


\abstract{Vortices are swirling regions of fluid that structure motion in gases and liquids across a wide range of scales, from laboratory-scale experiments to vast atmospheric currents. They play a key role in mixing, transport, and energy transfer, yet their reliable identification in unsteady flows remained a major challenge. Most existing approaches rely on local, instantaneous properties of the velocity gradient, such as strain or rotation. Although effective in simple or steady flows, these criteria can fail in complex, time-dependent settings, falsely detecting vortices or overlooking coherent structures altogether.\\
Lagrangian methods instead identify vortices as regions of material coherence by tracking fluid trajectories over time. While conceptually sound, these approaches are computationally intensive, require high-quality data, and are impractical for real-time applications. This motivates a central challenge: whether coherent vortices, inherently defined over finite times, can be detected objectively from instantaneous flow measurements.\\
Here we introduce the first Eulerian criterion to overcome these challenges. By examining the temporal evolution of strain and removing from the velocity field the components attributable to rigid-body motion, we construct an objective velocity field that isolates genuine swirling dynamics. The resulting $Q_\text{s}$-criterion consistently identifies coherent vortices in both analytical examples and complex flow data, including cases where traditional methods fail. Our framework provides an observer-independent, computationally efficient tool for vortex detection from instantaneous data, enabling improved analysis and prediction of fluid flows across scales.}

%%================================%%
%% Sample for structured abstract %%
%%================================%%




\maketitle

\section{Introduction}\label{Section 1}
Vortices - swirling, coherent regions of fluid motion - are a fundamental feature of many natural and engineered flows. They shape the dynamics of ocean currents that regulate climate \cite{Darryn2017}, influence the transport of nutrients and pollutants \cite{Lekien2008,Dauxois2021} such as microplastics \cite{BRACH2018191}, and determine the formation of storms and hurricanes \cite{Lorsolo2008} as well as mixing regions in the atmosphere \cite{Pierrehubert1993}. In engineering applications, vortices control the efficiency of wind turbines \cite{Stevens2017}, cause dynamic loading in fighter jets \cite{erickson1991windtunnel}, and find widespread application in chemical reactors \cite{SCZECHOWSKI19953163}. Given their ubiquity and importance, it is surprising that reliably identifying and characterizing vortices from instantaneous flow data still remains one of the central challenges in applied fluid mechanics \cite{Haller2021}. 

Indeed, classical approaches to vortex detection relying on instantaneous, or Eulerian velocity data, such as the Okubo--Weiss, $Q$-, $\lambda_2$-, $\Delta$-, and $\lambda_{\rm ci}$-criteria \citep{okubo1970horizontal,weiss1991dynamics,hunt1988eddies,jeong1995identification,chong1990general,chakraborty2005relationships}, measure rotation, shear, or strain to identify regions of coherent swirling motion. Most commonly, a vortex is identified via the symmetric and antisymmetric decomposition of the velocity gradient tensor $\nabla \bm{v} = \bm{S} + \bm{W}$, where $\bm{S}$ and $\bm{W}$ denote the rate-of-strain and spin tensors, respectively, leading to the $Q$-criterion
\begin{eqnarray}
Q &=& \frac{1}{2}\left( \|\bm{W}\|^2 - \|\bm{S}\|^2 \right),
\end{eqnarray}
with vortical regions defined by $Q>0$, where local rotation dominates over strain. By the same rationale, the $Q=0$ isosurface defines vortex boundaries. For planar incompressible flows, all the other, aforementioned criteria coincide with the $Q$-criterion. While these methods can be rigorously justified in incompressible, planar, and steady flows, they generally yield contradictory results in unsteady flows, sometimes falsely labeling regions as vortices or missing them entirely \citep{haller2005objective,Pedergnana20}. These deficiencies are directly linked to the lack of objectivity of the aforementioned criteria, leading to their dependence on observer changes of the form 
\begin{equation} \label{Euclidian frame change 2}
    \bm{x}=\bm{Q}(t)\bm{y}+\bm{b}(t),\quad \bm{Q}^T\bm{Q}=\bm{I}.
\end{equation}
The presence of a coherent vortex should, however, be consistently recognized by \textit{all} observers. Consequently, the lack of objectivity does not support the quantitative application of the $Q$-criterion - or any of its variations - to unsteady flows, which constitute the overwhelming majority of fluid flows of practical and geophysical relevance. These gaps in the understanding of fluid flows carry profound implications: even seemingly minor differences in the definition of vortices can cascade into substantial uncertainties in weather prediction, assessing pollutant transport, estimating energy capture, and understanding a host of geophysical and engineering flows. As a result, the reliable identification and characterization of vortices is not merely a technical challenge but a problem of fundamental scientific and practical importance.

\begin{figure}
    \centering
    \includegraphics[width=0.95\linewidth]{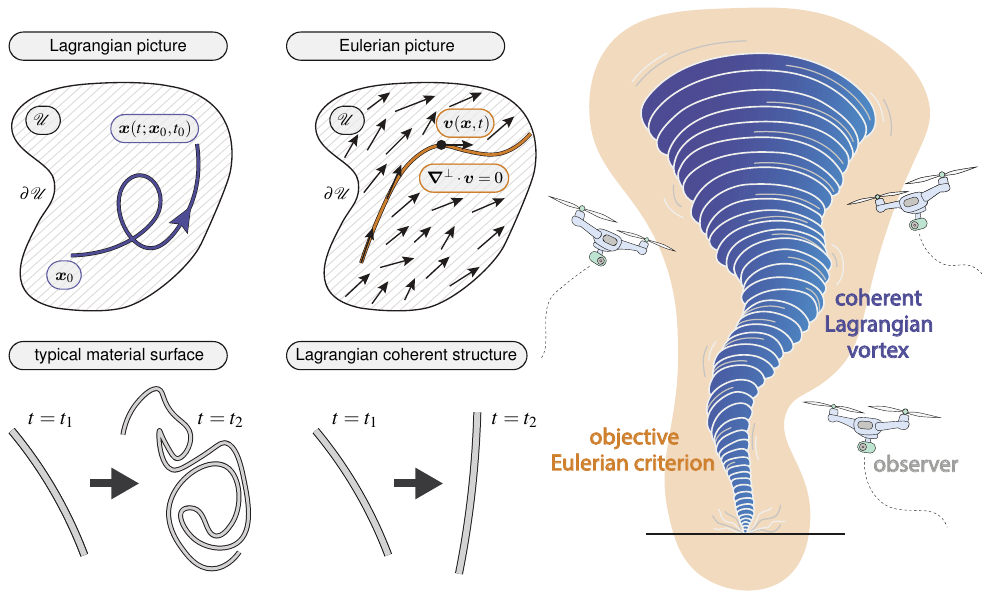}
    \caption{\textbf{Overview of different descriptions of fluid flows and their connection to coherent structures.} Fluid mechanics can be described in two different frameworks: The Lagrangian picture, in which particles are advected over time by the flow, and the Eulerian picture, which considers instantaneous velocity fields which are, at each point, tangential to streamlines $\bm{\nabla}^\perp\cdot \bm{v}=0$. In the Lagrangian picture, one can readily define coherent structures, i.e., collections of fluid particles which filament significantly less than their neighbors under advection over finite times. These flow surfaces are distinguished from other regions in the domain and play a crucial role in transport phenomena, mixing and vortex dynamics. The consistent detection of such Lagrangian coherent structures, especially of vortices, in an observer-independent (objective) fashion using Eulerian methods is an open challenge which is tackled in this work. }
    \label{Figure 1}
\end{figure}

In this work, we resolve this long-standing limitation by introducing an objective, physically grounded reformulation of Eulerian vortex detection for unsteady flows. By separating the flow field from the unsteady contribution of its instantaneous relative motion, described by the partial time derivative of the strain-rate tensor $\bm{S}$, we construct an observer-independent transformation of the instantaneous velocity field that filters out non-material, frame-dependent effects. This distinguished objective modification of the rate-of-strain tensor $\bm{S}$ is defined through the solution of a spatio-temporal variational principle. The objectively modified rate-of-strain tensor naturally leads to a generalized, objective vortex criterion, denoted $Q_\text{s}$, which extends the classical $Q$-criterion to unsteady flows while preserving its intuitive interpretation as a competition between rotation and strain. See Fig. \ref{Figure 1} for an illustration of the main result of this work.

Crucially, the proposed framework recovers the classical $Q$-criterion in steady, planar, incompressible flows - where all standard Eulerian criteria coincide - while remaining well-defined and physically meaningful in fully unsteady settings. As demonstrated through analytical benchmarks and numerical examples, $Q_\text{s}$ consistently identifies coherent vortical structures without spuriously classifying shear-dominated regions as vortices, and without missing material vortices that are invisible to traditional instantaneous criteria and their recent extensions \cite{Kaszás2023211,KogelbauerPedergnana2025}. The proposed method is the first and only instance of a vortex criterion in the literature that simultaneously (a) is objective, (b) resolves all known pathological examples of vortex identification, and (c) remains directly applicable to large, scale-rich data sets at a reasonable computational cost. While criteria satisfying (a) and (b) can, in principle, be constructed using trajectory-based Lagrangian methods \cite{haller2018material,Haller_2023}, such approaches are often impractical in realistic settings due to limitations in data quality, insufficient time steps for meaningful fluid particle advection, or the sheer size of the data sets involved. In summary, this work provides a rational and objective foundation for vortex identification in unsteady flows, combining physical consistency with practical applicability to large, complex data sets.

\section{Results}
Our main result is a modified, objective vortex criterion, which is invariant under observer changes of the form \eqref{Euclidian frame change}: 
\begin{equation}\label{Qs}
    Q_\text{s}(\bm{x},t) = \frac{1}{2}\big(\|{\bm{W}(\bm{x},t)-\operatorname{skew}{\left[\bm{T}(t)\right]}}\|^2-\|\bm{S}(\bm{x},t)-\operatorname{symm}{\left[\bm{T}(t)\right]}\|^2).
\end{equation}
According to this criterion, a vortex is defined as a region of fluid for which $Q_\text{s}>0$ and vortex boundaries are defined by level sets $Q_\text{s}=0$. A simple algorithm for computing $Q_\text{s}$ from the velocity field $\bm{v}(\bm{x},t)$ is given in the Methods section. The matrix $\bm{T}$ is defined in vectorized notation, see \cite{magnus2007matrix}, as 
\begin{eqnarray} \label{Linear equations}
    \operatorname{vec}[\bm{T}]
=\overline{\bm{M}^{T}
\bm{M}}^{-1}\,\overline{\bm{M}^{T}\operatorname{vec}[\partial_t \bm{S}]},
\end{eqnarray}
where the average $\overline{f}=\fint_\mathcal{D} f dV = \frac{1}{\mathrm{Vol}(\mathcal{D})}\int_{\mathcal{D}} f dV$ is taken over a certain domain $\mathcal{D}$,  $\bm{M} = \bm{S} \otimes \bm{I}
+ (\bm{I} \otimes \bm{S})\,\bm{K}$ is a linear function of the rate-of-strain tensor $\bm{S}$ and $\bm{K}$ satisfies $\operatorname{vec}[\bm{A}^T]=\bm{K}\,\operatorname{vec}[\bm{A}]$. An explicit expression for $\bm{M}$ in terms of the entries of $\bm{S}$ is given in the Methods section. The matrix $\bm{T}$ is defined as the solution of the spatio-temporal variational principle that seeks to minimize the action 
\begin{equation}\label{action}
    \mathcal{S}[\bm{T}]= \frac{1}{2} \int_{t_0}^{t_1} \fint_{\mathcal{D}} \|\overset{\circ}{\bm{S}}(\bm{x},t;\bm{T}(t))\|^2 dV dt,
\end{equation}
where $\overset{\circ}{\bm{S}}$ denotes a modified time derivative of the symmetric strain rate tensor $\bm{S}$:
\begin{eqnarray} \label{Objective rate}
    \overset{\circ}{\bm{S}}(\bm{x},t)=\partial_t \bm{S}(\bm{x},t)-\bm{T}(t)\bm{S}(\bm{x},t)-\bm{S}(\bm{x},t)\bm{T}^T(t).
\end{eqnarray}  
Details of the solution to the minimization of \eqref{action} are provided in the Supplementary Notes. Crucially, by the definition of $\bm{T}$, the $Q_\text{s}$-criterion, as well as the quantity $\overset{\circ}{\bm{S}}$ itself, are objective. As a consequence, \textit{all} observers identifying a vortex in region through $Q_\text{s}$ will be in agreement. Physically, the the matrix $\bm{T}$ minimizes the unsteady component of $\bm{S}$ through the modified time derivative and the variational principle \eqref{action}. See the Discussion section for a more detailed physical interpretation. If the domain is chosen sufficiently large, and the unsteady part of the flow is localized, the matrix $\bm{T}$ will be close to zero as it aims to compensate the unsteady contributions of $\bm{S}$ evenly over the domain. Thus, to appropriately capture the unsteadiness of the flow, the domain $\mathcal{D}$ has to be defined via an iterative procedure: First, $\overset{\circ}{\bm{S}}(\bm{x},t)$ is computed for $\bm{T}$ as defined in \eqref{Linear equations} on the whole flow domain. In a second step, the maximum $\bm{x}_\text{max}$ of $\|\overset{\circ}{\bm{S}}(\bm{x},t)\|$ at the given time step $t$ is computed. In a third step, $\mathcal{D}$ is chosen as a small cuboid centered at $\bm{x}_\text{max}$. This iterative procedure considerably improves the overall accuracy of $Q_{{\text{s}}}$, while retaining more details of the underlying flow field. The numerically specific choices of $\mathcal{D}$ used for the examples below are given in the Supplementary Notes. We now apply our criterion to three examples of unsteady flow data to demonstrate its prediction power.

\subsection{Objective vortex detection in unsteady flow data}\label{Example Section}
The first example describes a two-dimensional flow around a cylinder heated from below, see \cite{gunther2017generic}. This example was computed with the Gerris flow solver \cite{POPINET2003572} over $2000$ time steps using the Boussinesq approximation over the simulation domain $(x,y,t)\in [-0.5, 0.5] \times [-0.5, 2.5]\times [0, 20]$. The grid resolution is $150$ ($x$) $\times$ $450$ ($y$). To represent Lagrangian fluid-particle motion, and to serve as a reference measure of the coherent structures genuinely present in the flow, we compute the Finite-Time Lyapunov Exponent (FTLE) \cite{SHADDEN2005271}, which quantifies the exponential rate of separation of initially neighboring particle trajectories over finite time. Given the flow map $\boldsymbol{\Phi}_{t_0}^{t_0+T}:\boldsymbol{x}_0\mapsto \boldsymbol{x}(t_0+T;\boldsymbol{x}_0)$, which maps initial particle positions at time $t_0$ to their positions at time $t_0+T$, the FTLE field is defined as
\begin{equation}
\mathrm{FTLE}(\boldsymbol{x}_0,t_0,T)
=
\frac{1}{T}
\ln\sqrt{
\lambda_{\max}\!\left(
\mathbf{C}_{t_0}^{t_0+T}(\boldsymbol{x}_0)
\right)
},
\end{equation}
where $\lambda_{\max}$ denotes the largest eigenvalue of the Cauchy--Green strain tensor
\begin{equation}
\mathbf{C}_{t_0}^{t_0+T}
=
\left(\nabla \boldsymbol{\Phi}_{t_0}^{t_0+T}\right)^{T}
\nabla \boldsymbol{\Phi}_{t_0}^{t_0+T}.
\end{equation}
Ridges of maxima in the FTLE field are indicators of repelling LCSs in forward time ($T>0$) and of attracting LCSs in backward time ($T<0$) \cite{HALLER2001248}.

\begin{figure}
    \centering
    \includegraphics[width=1\linewidth]{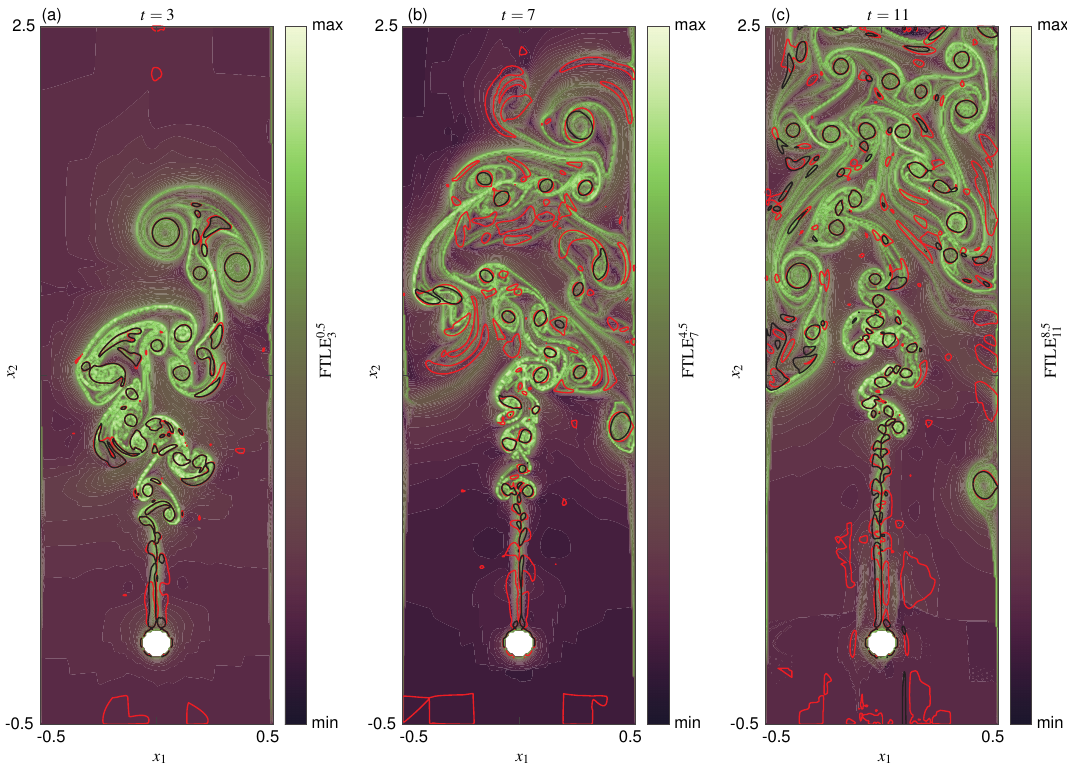}
    \caption{\textbf{Detecting coherent vortices in a two-dimensional flow around a cylinder.} The backward FTLE field (green lines) together with zero isocontours of the classical $Q$-criterion (red) and the objective $Q_\text{s}$-criterion (black). While the $Q$-criterion produces spurious and physically unrealistic vortex predictions, the $Q_{\text{s}}$-criterion correctly predicts Lagrangially coherent regions.}
    \label{Figure 2}
\end{figure}
\begin{figure}
    \centering
    \includegraphics[width=1\linewidth]{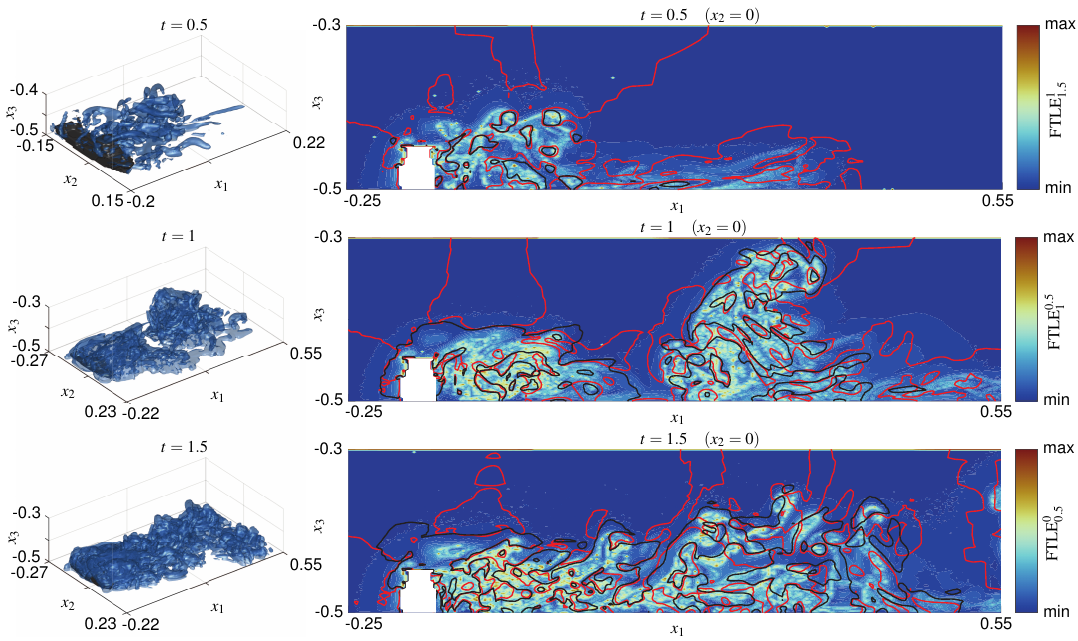}
    \caption{\textbf{Three-dimensional vortex analysis in the wind wake of the research vessel Tangaroa.} Left column: Objective three-dimensional zero-isocontours of the $Q_\text{s}$-criterion that align with Lagrangian coherent structures in the flow. Right column: Backward FTLE field over the mid-plane $(x_2=0)$ of the domain overlaid with zero isocontours of the classical $Q$-criterion (red) and the objective $Q_\text{s}$-criterion (black). While the $Q$-criterion produces spurious and physically unrealistic vortex predictions, the $Q_{\text{s}}$-criterion correctly predicts Lagrangially coherent regions.  }
    \label{Figure 3}
\end{figure}

In Fig. \ref{Figure 2}, we compare this Lagrangian measure (backward-time FTLE) to instantaneous predictions of vortices in the flow obtained using the $Q$-criterion (shown in red) and the $Q_\text{s}$-criterion (shown in black). While the FTLE field indicates vortices as spirals, and near-circular structures advected by the flow, we stress that there is no rigorous definition of a vortical structure from first principles in this flow example. The $Q$-and $Q_\text{s}$-criteria, on the other hand, define vortices simply by closed zero isocontours. In the three equally spaced time steps examined, the $Q$-criterion predicts numerous spurious features with no physical significance, and pronounces many features in the flow field as vortices which do not carry any Lagrangian significance. In contrast, the zero isocontours of $Q_\text{s}$ highlights true coherent structures to a large extent, which align well with ridges of maxima of the backward FTLE field.

As our second example, we analyze vortices in a simulated, unsteady three-dimensional flow data set describing the wind wake behind the Tangaroa research vessel operated by the National Institute of Water and Atmospheric Research of New Zealand \cite{Popinet04:Tangaroa}. The data set contains  the velocity field $\bm{v}(\bm{x},t)$ computed using the Gerris solver over 200 time steps over the domain $(x,y,z,t)\in[-0.35, 0.65] \times [-0.3, 0.3] \times [-0.5, -0.3] \times [0, 2]$. The grid resolution is $300$ ($x$) $\times$ $180$ ($y$) $\times$ $120$ ($z$). As in the example before, we compute the backward FTLE field over three time steps, see Fig. \ref{Figure 3} (right column) over the mid-plane ($x_2=0$) of the domain. Note that the FTLE field was computed by launching trajectories from the mid-plane and then advecting backwards over a time $T$ in the three-dimensional domain. In the left column, the zero isocontours of $Q_\text{s}$ in the three-dimensional space are shown in blue. In the right column, the zero isocontours in the mid-plane of the $Q$-and $Q_\text{s}$-criteria are shown are shown in red and black, respectively. We observe that while $Q_\text{s}$ aligns well with the Lagrangian coherent structures highlighted by the FTLE field, the $Q$-criterion again pronounces spurious features in the field which - for the most part - do not align with the coherent structures and are thus unrelated to the actual fluid particle motion defined by the velocity field. This example demonstrates that our method allows the production of succinct, observer-independent 3D isosurface contour images from instantaneous velocity data that align with the Lagrangian fluid particle motion. This motion would be, for example, observed in smoke visualization experiments \cite{10.1063/1.861721} or in the raw tracer images produced by laser particle image velocimetry (PIV) \cite{Faure-Beaulieu_Xiong_Pedergnana_Noiray_2023}.

As our third example we use a data set describing the dynamics of Hurricane Isabel from 1700 UTC 16 September until 1700 UTC 18 September 2003. The simulation covers a period of $48$ time steps (hours), where each time step contains the instantaneous velocity field $\bm{v}(\bm{x},t)$ over a domain ranging from $83.8$W to $62.8$W longitude, $23.78$N to $41.78$N latitude, and $0.035$km to $19.835$km altitude. We only consider the first few hours of this time span here in order to capture a the dominant part of its dynamics before the Hurricane weakened and made landfall on 18 September. The grid resolution is $500$ ($x$) $\times$ $500$ ($y$) $\times$ $100$ ($z$). We focus on the lowest altitude level ($35$m) and consider the data set as a purely two-dimensional data set on that specific altitude level. This approximately mimics real-time weather data, where wind speed is typically provided in a two-dimensional format at $10$m and, possibly, at an additional level in the low hundred meters. The European Centre for Medium-Range Weather Forecasts (ECMWF), for example, provides wind speeds at $10$m and $100$m. The classic Saffir--Simpson scale for the intensity of tropical hurricanes is also defined by the $10$m wind speed.

Since we are dealing with the first few time steps of a flow data set, there is no possibility to compute backward FTLE at over any significant time scale. In light of this limitation, we choose a proxy for Lagrangian coherence as a benchmark for $Q_{\text{s}}$ namely the total precipitation density, hereafter referred to as \textit{precipitation}. This quantity measures the mass content of graupel, rain and snow in the atmosphere, see \cite{isabel2004wrf}. Our rationale for considering this variable as a Lagrangian proxy is that coherent structures are known to be influential in the transport of inertial particles, such as precipitation \cite{InertialParticleDynamicsinaHurricane,Haller2016136}. The distribution of precipitation can thus be used as an indicator of the presence of Lagrangian coherent structures. 

In Fig. \ref{Figure 5}, we compare the zero isocontours of the $Q_\text{s}$-criterion (black) to those of the $Q$-criterion (red) against the precipitation over a subset of the full domain, roughly around the eye of the hurricane. The land mass is indicated in white. We observe that the $Q$-criterion pronounces countless spurious and physically meaningless structures, which grow worse and worse over time as the hurricane weakens and grows in size. In contrast, the $Q_\text{s}$-criterion describes isolated, compact regions that align astoundingly well with the precipitation data. The last time step, where $t=t_0+8h$, is particularly noteworthy, as it shows near-perfect alignment of the $Q_\text{s}$-criterion with the precipitation data and even with the rainband \cite{barnes83} in the far-out southeast part of the domain. We emphasize that the side length of each square in the figure is roughly $1200$km long, implying that the closed contours, which appear small in the figure, represent kilometer-sized vortices, predicted from instantaneous velocity data, that align with Lagrangian coherent structures in the flow.

\begin{figure}[t]
    \centering
    \includegraphics[width=1\linewidth]{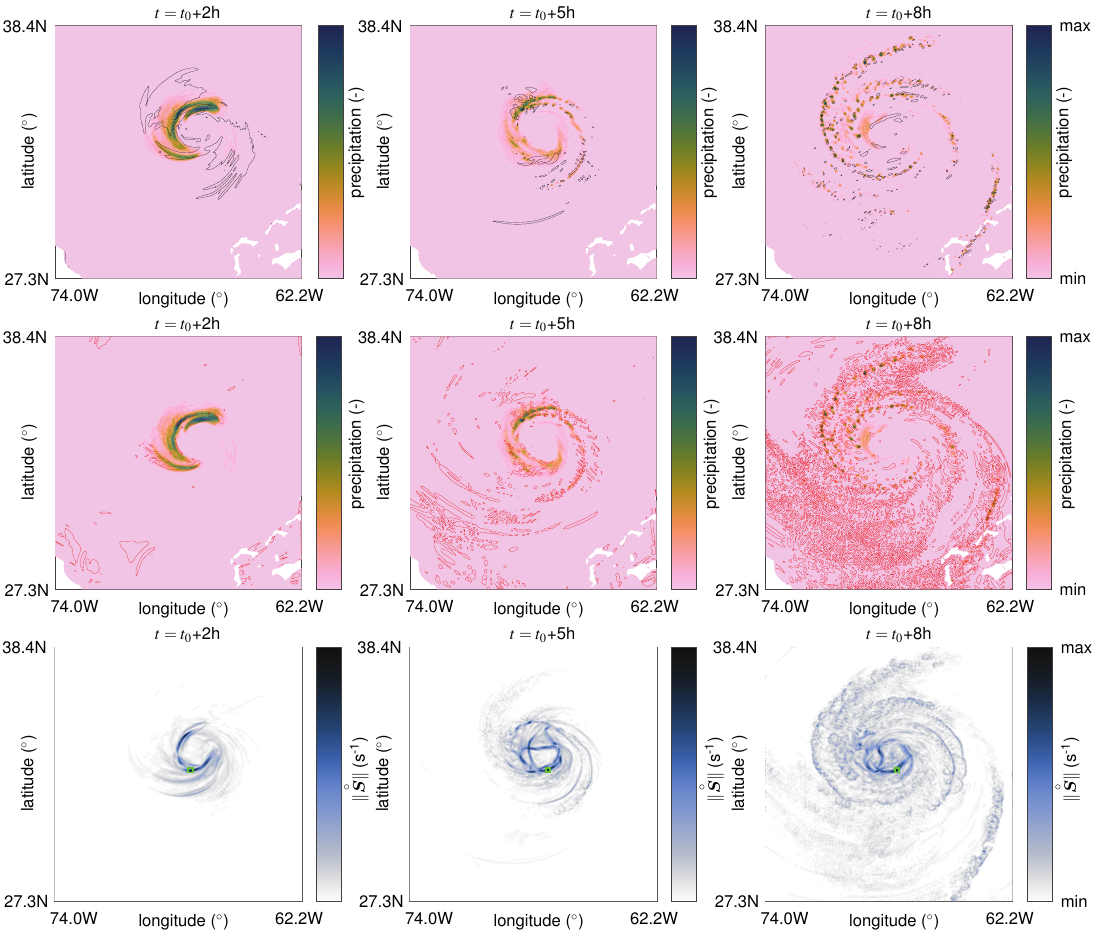}
    \caption{\textbf{Prediction of vortices from instantaneous velocity data of the Hurricane Isabel.} First and second row: Precipitation overlaid with zero isocontours of the $Q_\text{s}$ (balck) and $Q$-criteria (red). Third row: Contour plot of the normed objective modified rate $\|\overset{\circ}{\bm{S}}\|$. All data is shown for three equally space time steps.}
    \label{Figure 4}
\end{figure}

The bottom row in Fig. \ref{Figure 6} shows a contour plot of the objective rate $\|\overset{\circ}{\bm{S}}\|$, visualizing the spread of the hurricane as well as the small domains $\mathcal{D}$ (green squares) which were chosen to compute $\bm{T}$ for each time step, see \eqref{Linear equations}. These contour plots indicate an intricate pattern of near-circular tubes that appear to be knotted together, reminiscent of the vortex tube dynamics observed in laboratory-scale experiments \cite{Scheeler2017}. The analysis of the hurricane data presented here evidences $Q_{\text{s}}$ as a powerful tool for the analysis of weather data. This further supports the potential of $Q_{\text{s}}$ as a predictor of key Lagrangian features, such as transport of moisture and similar tracers of fluid particle motion.

\section*{Methods}
\label{Methods}
\setcounter{subsection}{0}
\subsection*{Vortex detection algorithm}
A complete description of the objective vortex detection algorithm presented in this work is given in the pseudocode below. For simplicity of presentation, we consider a single domain, which is then incorporated into an iterative procedure in which $\bm{T}$ is first computed over the whole flow domain. Then, after finding the maxima of the modified time derivative $\|\overset{\circ}{\bm{S}}(\bm{x},t)\|$ for each time step, the matrix $\bm{T}$ is computed again over over a small cuboid around the identified maxima.

{\centering
\setlength{\FrameRule}{0.8pt}
\setlength{\FrameSep}{8pt}
\begin{framed}
\noindent\textbf{Objective $Q_\text{s}$-Criterion Computation}\\ \vspace{0.2cm}

\flushleft\textbf{Input:} Velocity field $\bm{v}(\bm{x},t)$ on a 3D domain $\mathcal{D}$

\textbf{Output:} Objective, instantaneous vortex boundaries defined by $Q_\text{s}(\bm{x},t)=0$

\begin{enumerate}
    \item Compute the strain-rate tensor:
    $$\bm{S}(\bm{x},t) = \frac{1}{2} \left( \nabla \bm{v} + \nabla \bm{v}^T \right)$$
    \item Construct the Kronecker-based matrix:
    $$\bm{M}(\bm{x},t) = \bm{S} \otimes \bm{I} + (\bm{I} \otimes \bm{S}) \bm{K},$$
    where $\bm{K}$ satisfies $\operatorname{vec}[\bm{A}^T] = \bm{K}\,\operatorname{vec}[\bm{A}]$.
    
    \item Solve the spatio-temporal minimization problem for $\bm{T}(t)$:
    $$\operatorname{vec}[\bm{T}] = \overline{\bm{M}^T \bm{M}}^{-1} \;\overline{\bm{M}^T \operatorname{vec}[\partial_t \bm{S}]}.$$
    Or, if $\overline{\bm{M}^T \bm{M}}$ is singular, use the Moore--Penrose pseudoinverse:
    $$\operatorname{vec}[\bm{T}] = \overline{\bm{M}^T \bm{M}}^{+} \;\overline{\bm{M}^T \operatorname{vec}[\partial_t \bm{S}]},$$
    corresponding to the least-squares solution of the minimization problem.
    %\item Define the auxiliary velocity field $\bm{v}_\text{s}$:$$\bm{v}_\text{s}(\bm{x},t) = \bm{v}(\bm{x},t) - \bm{T}(t)\big[\bm{x}-\overline{\bm{x}}(t)\big] - \overline{\bm{v}}(t),$$where $\overline{\bm{x}}$ and $\overline{\bm{v}}$ are the spatial averages of position and velocity.
    \item Compute the objective $Q$-criterion:
    $$Q_\text{s} = \frac{1}{2} \Big( \|\bm{W}-\operatorname{skew}[\bm{T}]\|^2 - \|\bm{S}-\operatorname{symm}[\bm{T}]\|^2 \Big),$$
    where $\bm{W} = \frac{1}{2}(\nabla \bm{v} - \nabla \bm{v}^T)$.
    
    \item Extract vortical regions using the threshold $Q_\text{s} = 0$.
\end{enumerate}
\end{framed}}

The following formulae may be useful for the practical implementation of the algorith above: For two-dimensional flows, the matrix $\bm{M}$ is given, in terms of the entries $S_{ij}$, $i,j\in \{1,2\}$ of the rate-of-strain tensor $\bm{S}$, by
\begin{eqnarray}
    \bm{M}=\begin{pmatrix}
2S_{11} & 0      & 2S_{12} & 0 \\
S_{12}  & S_{11} & S_{22}  & S_{12} \\
S_{12}  & S_{11} & S_{22}  & S_{12} \\
0       & 2S_{12}& 0       & 2S_{22}
\end{pmatrix}.
\end{eqnarray}
For three-dimensional flows, $\bm{M}$ is given by 
\begin{eqnarray}
    \bm{M}=\begin{pmatrix}
2S_{11} & 0      & 0      & 2S_{12} & 0      & 0      & 2S_{13} & 0      & 0 \\
S_{12}  & S_{11} & 0      & S_{22}  & S_{12} & 0      & S_{23}  & S_{13} & 0 \\
S_{13}  & 0      & S_{11} & S_{23}  & 0      & S_{12} & S_{33}  & 0      & S_{13} \\
S_{12}  & S_{11} & 0      & S_{22}  & S_{12} & 0      & S_{23}  & S_{13} & 0 \\
0       & 2S_{12}& 0      & 0       & 2S_{22}& 0      & 0       & 2S_{23}& 0 \\
0       & S_{13} & S_{12} & 0       & S_{23} & S_{22} & 0       & S_{33} & S_{23} \\
S_{13}  & 0      & S_{11} & S_{23}  & 0      & S_{12} & S_{33}  & 0      & S_{13} \\
0       & S_{13} & S_{12} & 0       & S_{23} & S_{22} & 0       & S_{33} & S_{23} \\
0       & 0      & 2S_{13}& 0       & 0      & 2S_{23}& 0       & 0      & 2S_{33}
\end{pmatrix}.
\end{eqnarray}
For completeness, the iterative procedure used in the Results section is also described below. Note that, by the objectivity of the modified rate $\overset{\circ}{\bm{S}}$, the location of its maximum is also objective. Hence, the algorithm shown below describes an objective procedure to enhance vortex detection using the $Q_\text{s}$-criterion. %As mentioned above in the Results section, this iterative procedure was informed by the analytical benchmarks presented below in the discussion section. 
The choice of the side lengths of the cuboid $\epsilon_{x,y,z}$ is user-dependent. As a rule of thumb, we suggest choosing the side lengths at a few percent of the respective grid axes. Note that the magnitude of $\bm{T}$ will typically tend to increase as the $\epsilon_{x,y,z}$ are  decreased. See the Discussion section below for a physical interpretation of this method and for a motivation of the iterative procedure from analytical benchmarks. To obtain a further improvement of the method, the side lengths of the reduced domain could be adapted at each time step. This additional layer of refinement requires another metric to decide whether the domain should be shrunk or grown, which may be a fruitful topic to tackle with machine-learning methods.

{\centering
\setlength{\FrameRule}{0.8pt}
\setlength{\FrameSep}{8pt}
\begin{framed}
\noindent\textbf{Iterative procedure for enhanced vortex detection}\\ \vspace{0.2cm}

\flushleft\textbf{Input:} Velocity field $\bm{v}(\bm{x},t)$ on a 3D domain $\mathcal{D}$

\textbf{Output:} Objective, instantaneous vortex boundaries defined by $Q_\text{s}(\bm{x},t)=0$

\begin{enumerate}
    \item Using the method described above in ``Objective $Q_\text{s}$-Criterion Computation'', compute the matrix $\bm{T}$ over the whole flow domain.
    \item Compute the matrix norm of the modified rate, $$\|\overset{\circ}{\bm{S}}\|=\|\partial_t \bm{S}(\bm{x},t)-\bm{T}(t)\bm{S}(\bm{x},t)-\bm{S}(\bm{x},t)\bm{T}^T(t)\|,$$ over the full domain and find the location of its maximum $\bm{x}_\text{max}$.
    \item Define a new domain as a cuboid centered at $\bm{x}_\text{max}$ with side lengths $\epsilon_{x,y,z}$.
    \item Repeat the method described above in ``Objective $Q_\text{s}$-Criterion Computation'' for the domain inside the cuboid and obtain $\bm{T}$ in this manner.
    \item Compute the objective $Q$-criterion using the newly obtained $\bm{T}$:
    $$Q_\text{s} = \frac{1}{2} \Big( \|\bm{W}-\operatorname{skew}[\bm{T}]\|^2 - \|\bm{S}-\operatorname{symm}[\bm{T}]\|^2 \Big),$$
    where $\bm{W} = \frac{1}{2}(\nabla \bm{v} - \nabla \bm{v}^T)$.
    \item Extract vortical regions using the threshold $Q_\text{s} = 0$.
\end{enumerate}
\end{framed}}

\section*{Discussion}

%\textcolor{red}{Discussion section normalerweise eine zusammenfassung der resultate, weitere objectives etc oder? Ist das so eine Format-Vorgabe von Nat. Comm.?}
%\textcolor{blue}{Ich habe es so gelernt: Es gibt keine zusammenfassende Conclusion-type Diskussion der Resultate in Nature. Alle Diskussion der Resultate ist generell in der Results section selber. Die Methods und Discussion Sections sind für technische und weiterführende Details. Es wäre natürlich super, wenn Du am z.B. am Ende der Results section noch etwas zu Anwendungen, Geoengineering, Rainbands etc. hinzufügen könntest.}

\subsection*{Domain-dependence}
The spectrum of the matrix $\bm{M}(\bm{x},t)\in \mathbb{R}^{d^2\times d^2}$ as defined after \eqref{Linear equations} is given by
\begin{eqnarray}
\operatorname{Spec}(\bm{M}) &=& \{0\} \cup \{ \lambda_i + \lambda_j \,:\, i,j = 1,\dots,d \}, 
\end{eqnarray}
where $\lambda_i$, $i=1,\dots,d$ are the eigenvalues of the rate-of-strain tensor $\bm{S}(\bm{x},t)$, see \cite{magnus2007matrix}. In particular, the matrix $\bm{M}$ is singular. We recall that the singularity of a matrix acting in the space of vectorized matrices is not to be confused with a singular matrix in ordinary vector space. Indeed, a singular matrix $\bm{A}$ acting on a vectorized matrix $\operatorname{vec}\left({B}\right)$ produces a vectorized matrix $\operatorname{vec}\left({C}\right)=\bm{A}\operatorname{vec}\left({B}\right)$, where $\bm{C}$ is nonsingular in ordinary vector space in general. In other words, the singularity of $\bm{M}$ is simply an artifact of matrix vectorization.

Despite $\bm{M}$ being pointwise singular at all $(\bm{x},t)$, its spatial average $\overline{\bm{M}^{T}\bm{M}}$ is generally nonsingular. Indeed, if the domain is chosen such that contains more than a single grid point over which the spatial average $\overline{\bm{M}^{T}\bm{M}}$ is computed, singularity of $\overline{\bm{M}^{T}\bm{M}}$ would be equivalent to exact numerical cancellation of all the spatially varying eigenvectors of $\bm{S}$. If this very degenerate case occurs, one slightly changes the domain or adopts the least-squares solution of the minimization problem. 

To illustrate this singularity further, consider a special analytical solution where the unsteady part of the velocity field $\bm{v}$ is linear in $\bm{x}$. In this case, $\bm{M}$ is constant and the spatial average of $\overline{\bm{M}^{T}\bm{M}}$ is thus singular. The least-squares solution to the problem of minimizing the action \eqref{action} can the be obtained as
\begin{eqnarray} \label{Linear equations pseudo}
    \operatorname{vec}[\bm{T}_\text{lsq}]
=\overline{\bm{M}^{T}
\bm{M}}^{+}\,\overline{\bm{M}^{T}\operatorname{vec}[\partial_t \bm{S}]}
\end{eqnarray}
where $\bm{A}^{+}$ denotes the Moore--Penrose pseudoinverse $\overline{\bm{M}^{T}
\bm{M}}^{+}$, see \cite{magnus2007matrix}. The matrix $\bm{T}_\text{lsq}$ defined by \eqref{Linear equations pseudo} transforms like a spin tensor if
\begin{eqnarray} \label{condition}
    (\bm{Q}^T \!\otimes\! \bm{Q}^T)\bm{M}^+\bm{M}\,\operatorname{vec}[\dot{\bm{Q}} \bm{Q}^T]=\operatorname{vec}[ \bm{Q}^T\dot{\bm{Q}} ], 
\end{eqnarray}
see the Supplementary Notes.

We stress that the use of the pseudoinverse in \eqref{Linear equations pseudo} does not necessarily imply that this solution is only approximate. Indeed, we discuss below a spatially linear flow field where $\overline{\bm{M}^{T}\bm{M}}$ is singular but the exact global minimizer of \eqref{action} can be directly obtained via \eqref{Linear equations pseudo}.

\subsection*{Objectivity of $Q_\text{s}$} 
In this work, we consider the Eulerian position vector $\bm{x}\in\mathcal{D}\subset\mathbb{R}^d$, $d=2,3$, defined on a time-independent, connected domain $\mathcal{D}$ and a sufficiently smooth, unsteady vector field $\bm{v}:\mathbb{R}^d\times [0,\infty)\to\mathbb{R}^d$. For simplicity, we exclude unbounded domains or domains whose boundary changes with time although these cases could be treated in a similar way. 
For any function $f:\mathcal{D}\to \mathbb{R}$, we define its spatial average as  
\begin{equation}
    \overline{f}=\fint_\mathcal{D} f dV = \frac{1}{\mathrm{Vol}(\mathcal{D})}\int_{\mathcal{D}} f dV.
\end{equation}
A general frame change in the Eulerian picture is described by the following transformations:
\begin{equation} \label{Euclidian frame change}
    \bm{x}=\bm{Q}(t)\bm{y}+\bm{b}(t),\quad \bm{Q}^T\bm{Q}=\bm{I},\quad \bm{Q}\in SO(\mathbb{R}^d),\quad \bm{b}\in \mathbb{R}^d.
\end{equation}
A quantity is called \textit{objective} if it transforms neutrally under a frame change of the form \eqref{Euclidian frame change}, see \cite{TruesdellNoll2004}. More specifically, an Eulerian scalar $\alpha$, a vector $\bm{a}$ or a matrix $\bm{A}$ is called \textit{objective} if it transforms according to 
\begin{equation}
    \tilde{\alpha}(\bm{y},t)=\alpha(\bm{x},t) \quad \text{or} \quad \tilde{\bm{a}}(\bm{y},t)=\bm{Q}^T(t){\bm{a}}(\bm{x},t)\quad \text{or} \quad  \tilde{\bm{A}}(\bm{y},t)=\bm{Q}^T(t){\bm{A}}(\bm{x},t) \bm{Q}(t),
\end{equation}
in the $\bm{y}$-frame induced by \eqref{Euclidian frame change}. Recall that the velocity field itself is non-objective since it transforms according to
\begin{equation}
    \tilde{\bm{v}}(\bm{y},t)=\bm{Q}^T(t)[{\bm{v}}(\bm{x},t)-\dot{\bm{Q}}(t)\bm{y}-\dot{\bm{b}}].
\end{equation}
Throughout this work, we denote quantities in the transformed frame with a tilde and the coordinates in the transformed frame are denoted by $\bm{y}$ as defined in Eq. \eqref{Euclidian frame change}. 

In the Supplementary Notes, we show that the matrix $\bm{T}$ transforms like a spin tensor under a general frame change:
\begin{eqnarray} \label{transformation property}
    \widetilde{\bm{T}}=\bm{Q}^T \bm{T} \bm{Q}-\bm{Q}^T \dot{\bm{Q}}.
\end{eqnarray}
We emphasize that this transformation property only holds for the minimizer of \eqref{action}, and is not an \textit{a priori} invariance of the action. As a consequence of \eqref{transformation property}, the modified time-derivative \eqref{Objective rate} is objective for the minimizers $\bm{T}$:
\begin{eqnarray}
\widetilde{\overset{\circ}{\bm{S}}}&=&\bm{Q}^T\partial_t \bm{S} \bm{Q}+\dot{\bm{Q}}^T\bm{S}\bm{Q}+\bm{Q}^T\bm{S}\dot{\bm{Q}}-\widetilde{\bm{T}}\bm{Q}^T{\bm{S}}\bm{Q}-\bm{Q}^T{\bm{S}}\bm{Q}\widetilde{\bm{T}}^T \notag\\
&=&\bm{Q}^T\left(\partial_t \bm{S}-\dot{\bm{Q}}\bm{Q}^T\bm{S}-\bm{S}\bm{Q}\dot{\bm{Q}}^T\right)\bm{Q}-\widetilde{\bm{T}}\bm{Q}^T{\bm{S}}\bm{Q}-\bm{Q}^T{\bm{S}}\bm{Q}\widetilde{\bm{T}}^T \notag\\
&=&\bm{Q}^T\left(\partial_t \bm{S}-{\bm{T}}{\bm{S}}-{\bm{S}}{\bm{T}}^T\right)\bm{Q}.
\end{eqnarray}
where we used the property that {$\dot{\bm{Q}} = -\bm{Q}\dot{\bm{Q}}^T\bm{Q}$} for the time-derivative of rotation matrices. Furthermore, it implies that the auxiliary velocity field
\begin{eqnarray} \label{objective velocity field}
    \bm{v}_\text{s}(\bm{x},t)=\bm{v}(\bm{x},t)-\bm{T}(t) \left[\bm{x}-\overline{\bm{x}}(t)\right]-\overline{\bm{v}}(t),
\end{eqnarray}
is objective. In Eq. \eqref{objective velocity field}, $\overline{\bm{x}}(t)$ and $\overline{\bm{v}}(t)$ are the location and velocity of the domain's center. We can thus introduce the following objective vortex criterion by computing the $Q$-criterion for $\bm{v}_\text{s}$:
\begin{equation}\label{Q}
    Q_\text{s} = \frac{1}{2}\big(\|{\bm{W}(\bm{x},t)-\operatorname{skew}{\left[\bm{T}(t)\right]}}\|^2-\|\bm{S}(\bm{x},t)-\operatorname{symm}{\left[\bm{T}(t)\right]}\|^2).
\end{equation}
In this definition, $\operatorname{skew}{\left[\bm{T}\right]}$ and $\operatorname{symm}{\left[\bm{T}\right]}$ are the skew-symmetric and symmetric parts of $\bm{T}$.

\subsection*{Physical interpretation}
Physically, the method presented in this work can be interpreted as follows: Through the action of the functional \eqref{Objective rate}, the matrix $\bm{T}$ compensates part of the unsteadiness of the rate-of-strain tensor $\partial_t \bm{S}$. This leads to the modified time derivative $\overset{\circ}{\bm{S}}\approx \bm{0}$, which, through the correction with $\bm{T}$, is closer to being zero in mean. Defining the objective auxiliary velocity field
\begin{eqnarray} 
    \bm{v}_\text{s}(\bm{x},t)=\bm{v}(\bm{x},t)-\bm{T}(t) \left[\bm{x}-\overline{\bm{x}}(t)\right]-\overline{\bm{v}}(t),
\end{eqnarray}
we note that the modified rate $\overset{\circ}{\bm{S}}$ for $\bm{v}_\text{s}$ satisfies 
\begin{eqnarray} \label{Objective rate vs}
    \overset{\circ}{\bm{S}}\left[{\bm{v}_\text{s}}\right](\bm{x},t)\approx -\bm{T}(t).
\end{eqnarray}
In words, we seek a matrix $\bm{T}$ such that the rate-of-strain tensor of the objective velocity field $\bm{v}_\text{s}$ has spatially uniform unsteadiness. By its uniformity, this unsteadiness can not perturb the predictions of local vortex criteria which are based on instantaneous, infinitesimal motion. Indeed, $Q_\text{s}$ is then simply the classical $Q$-criterion computed for $\bm{v}_\text{s}$.

\subsection*{Locations of maxima of $\|\overset{\circ}{\bm{S}}\|$}
\begin{figure}
    \centering
    \includegraphics[width=1\linewidth]{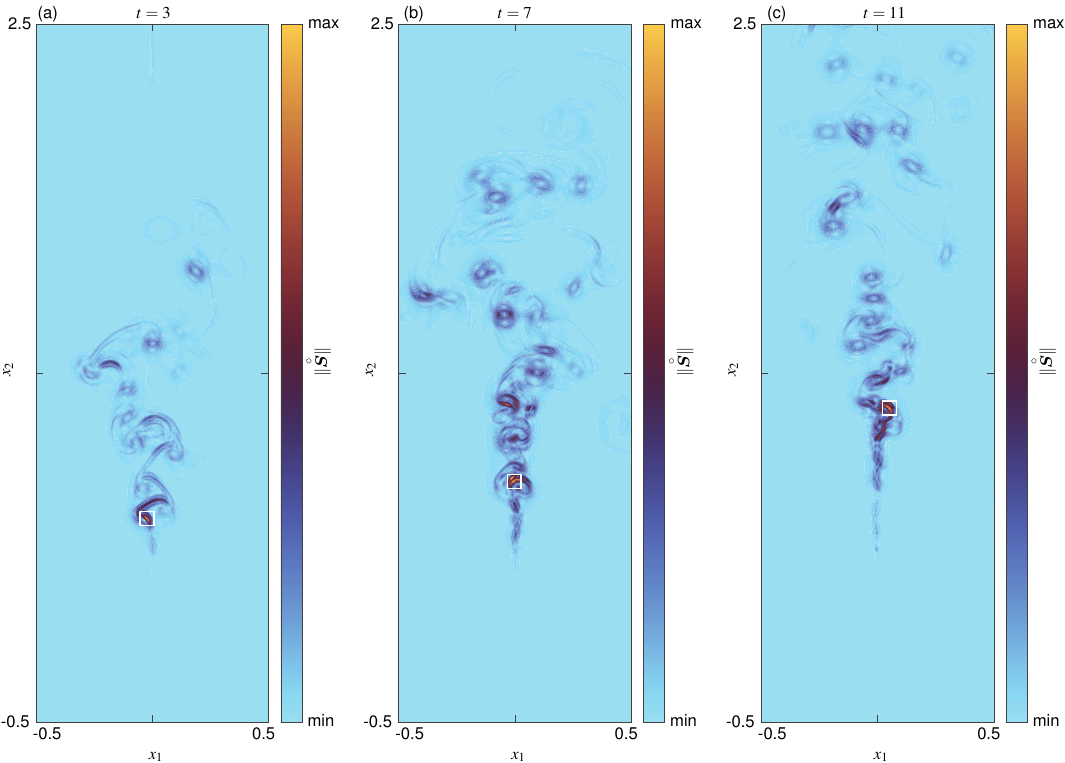}
    \caption{Cuboid domains (white) around the maxima of $\|\overset{\circ}{\bm{S}}\|$ for the example shown in Fig. \eqref{Figure 2}. These domains were used in the iterative procedure for enhanced vortex identification described in the Methods section.}
    \label{Figure 6}
\end{figure}

The small cuboid domains around the maxima of $\|\overset{\circ}{\bm{S}}\|$ used in the iterative procedure for enhanced vortex identification described in the Methods section are discussed here in more detail. For the first example in Section \ref{Section 1}, a heated flow around a cylinder, the cuboids are shown in Fig. \ref{Figure 6}. The maxima for the second example, the wind wake behind the research vessel Tangaroa, are shown in Table \ref{Table 1}. The numerical values of $\epsilon_k$, $k=x,y,z$, in that example were $\epsilon_x=0.04$, $\epsilon_y=0.01$, $\epsilon_z=0.04$. For the third example, Hurricane Isabel, the cuboid domains are shown in the bottom row of Fig. \eqref{Figure 5}.
\begin{table}[h!]
\centering
\begin{tabular}{|c|c c c |}
\hline
$t$ & $x_\text{max}$ & $y_\text{max}$ & $z_\text{max}$ \\
\hline
0.5  & $-0.0256$ & $-0.0486$ & $-0.4513$ \\
1  &  $0.2219$ & $0.0486$ & $-0.4395$ \\
1.5   &  $0.0747$ & $-0.1056$ & $-0.5000$ \\
\hline
\end{tabular}
\caption{Locations of maxima of $\|\overset{\circ}{\bm{S}}\|$ for the example shown in Fig. \eqref{Figure 4}. The side lengths of the corresponding cuboid domain $\epsilon_k$, $k=x,y,z$, in that example were $\epsilon_x=0.04$, $\epsilon_y=0.01$, $\epsilon_z=0.04$. The corresponding domains were used in the iterative procedure for enhanced vortex identification described in the Methods section.}
\label{Table 1}
\end{table}

\subsection*{Analytical benchmarks}
We now examine four different analytical examples, in all of which the true Lagrangian motion is known explicitly. On the one hand, these benchmarks serve to test and, on the other hand, inform the method used in this work. In the first two of these analytical examples (``Spatially linear Navier--Stokes flow'' and ``Separation and reattachment flow'') computing $\bm{T}$ over the full domain already gives the correct results and there is no need to adapt the domains. In the third and fourth examples domain-dependence plays a role. The polynomial nature of the spatial nonlinearity exhibited by third example  (``Nonlinear Navier--Stokes flow'') helps to clarify the dependence of the variational principle on the domain.  The last example in this section (``Four centers flow''), albeit being purely kinematic, has a more realistic flow geometry due to the relevant region being highly localized in space.

\subsubsection*{Spatially linear Navier--Stokes flow \label{Linear field example 1}} 
Consider the spatially linear family of unsteady Navier--Stokes solutions given by \cite{Pedergnana20}
\begin{eqnarray}
    \bm{v}(\bm{x},t)=\bm{A}(t)\bm{x}, \quad \bm{A}(t)= \begin{pmatrix}
    -\sin(Ct) & \cos(Ct) - \frac{{\omega}}{2}  \\
    \cos(Ct) + \frac{{\omega}}{2} & \sin(Ct) 
\end{pmatrix}.  \label{Linear field}
\end{eqnarray}
over the square domain $[-L/2,L/2]^2$. This flow is a generalization of the classic example introduced by Haller \cite{haller2005objective} in his seminal study on the shortcomings of Galilean invariant vortex criteria. In this example, $\bm{M}$ does not depend on $\bm{x}$ and $\overline{\bm{M}^{T}\bm{M}}$ is singular with rank three. We therefore apply the least-squares solution formula \eqref{Linear equations pseudo}. The condition \eqref{condition} is satisfied in this example for general $\bm{Q}\in SO(2)$, implying that the least-squares solution \eqref{Linear equations pseudo} is objective.

Since the field \eqref{Linear field} is spatially linear, any subset of the real plane can be chosen to take the spatial averages. From Eq. \eqref{Linear equations pseudo}, we obtain
\begin{equation}
    \bm{T}_\text{lsq}= \begin{pmatrix}
    0 & -\frac{C}{2}\\ 
    \frac{C}{2} & 0 
\end{pmatrix}. \label{matrix T}
\end{equation}
The modified rate $\overset{\circ}{\bm{S}}$ vanishes identically in this example, showing that $\bm{T}_\text{lsq}$ is the exact global minimizer of \eqref{action}. The objective SUV field $\bm{v}_\text{s}$ defined in \eqref{objective velocity field} is given by
\begin{eqnarray}
    \bm{v}_{\text{s}}(\bm{x},t)&=&\left[\bm{A}(t)-\bm{T}_{\text{lsq}}\right]\bm{x} = \begin{pmatrix}
    -\sin(Ct) & \cos(Ct) - \frac{{\omega-C}}{2} \\
    \cos(Ct) + \frac{{\omega-C}}{2} & \sin(Ct) 
\end{pmatrix}\bm{x}. \label{Objective Linear Field}
\end{eqnarray}
The velocity gradient of $\bm{v}_\text{s}$ is given by $\bm{\nabla} \bm{v}_{\text{s}}=\bm{A}(t)-\bm{T}$ with eigenvalues 
\begin{eqnarray}
    \lambda_{\pm}=\pm\dfrac{\sqrt{1-(\omega-C)^2}}{2}, \label{A evs}
\end{eqnarray}
predicting a vortical flow around the origin for $\vert \omega - C \vert > 2$ and a shear flow for $\vert\omega - C\vert < 2$. Equivalent predictions are obtained by the objective $Q_\text{s}$-criterion defined in \eqref{Q}. These parameter ranges correspond exactly to the fluid particle motion \cite{Pedergnana20}, see Fig. \ref{Figure 5}(a), which is judged incorrectly by the classical $Q$-criterion and its objective analogues $Q_\text{RB}$ and $Q_\text{US}$. See Fig. \ref{Figure 5}(b)-(d) for the respective predictions of the flow type by the other vortex criteria. We emphasize that no vortex criterion up to now has managed to consistently and correctly judge the fluid particle motion of this simple example, which has been introduced over two decades ago.

\subsection*{Separation and reattachment flow \label{separation flow example}}
We now consider the kinematic example of \cite{Lekien2008} given by the unsteady stream function defined on the unit circle, see also \cite{Kaszás2023211}:
\begin{eqnarray}
    \psi(\bm{x}, t) = \left( \vert\bm{x}\vert^2 - 1 \right)\left[ x_1 \sin(\omega_s t) + x_2 \cos(\omega_s t) \right] 
- \tfrac{1}{2}\,\omega_s \vert\bm{x}\vert^2.\label{streamfunction}
\end{eqnarray}
The matrix $\overline{\bm{M}^{T}\bm{M}}$ is nonsingular for all $t$. Computing the solution \eqref{Linear equations} over the whole flow domain  - the unit circle - yields 
\begin{eqnarray}
    \bm{T}= \begin{pmatrix}
    0 & \frac{\omega_s}{2}\\ 
    -\frac{\omega_s}{2} & 0 
\end{pmatrix}. \label{matrix T ex 2}
\end{eqnarray}
In this example, the modified rate \eqref{Objective rate} vanishes identically as well, showing that $\bm{T}$ is the exact global minimizer. The SUV field $\bm{v}_\text{s}=\bm{v}-\bm{T}\bm{x}$ corresponds to the streamfunction
\begin{eqnarray}
    \psi_\text{s}(\bm{x}, t) = \left( \vert\bm{x}\vert^2 - 1 \right)\left[ x_1 \sin(\omega_s t) + x_2 \cos(\omega_s t) \right].
    \label{streamfunction2}
\end{eqnarray}
In contrast to the original stream function, $\psi_s$ correctly predicts the topology of the fluid particle motion defined by \eqref{streamfunction}, which consists of two diametrically opposed vortex cells separated by a rotating separation line, see Fig. \ref{Figure 1}\textbf{b},\textbf{c}. This example has been previously resolved \cite{Kaszás2023211,KogelbauerPedergnana2025} and serves only as a check that the method described in this work measured up to these previous efforts for this kinematic example of an unsteady flow.

\subsubsection*{Nonlinear Navier--Stokes flow}
We use the results derived in this work to examine the spatially cubic Navier--Stokes field presented in \cite{Pedergnana20} given by
\begin{eqnarray} \label{unsteady 2d NS}
    \bm{u}(\bm{x},t)=\begin{pmatrix}
\sin(4t) & \cos(4t)+2 \\
\cos(4t)-2 & -\sin(4t)
\end{pmatrix}
\bm{x}\;+\;\alpha(t)
\begin{pmatrix}
x\,(x^{2}-3y^{2}) \\
-\,y\,(3x^{2}-y^{2}),
\end{pmatrix}
\end{eqnarray}
over the square domain $[-L,L]^2$. For simplicity, we consider here the special case where $\alpha$ is a real constant with a small value, so that the nonlinear flow defined by \eqref{unsteady 2d NS} retains the flow type of the spatially linear unsteady field \eqref{Linear field}, see \cite{Pedergnana20}. The matrix $\overline{\bm{M}^T\bm{M}}$ is nonsingular for all $t$, and $\bm{T}$ is given by
\begin{eqnarray}
    \bm{T}=\begin{pmatrix}
0 & \dfrac{40}{7L^{4}\alpha^{2}+20}  \\[6pt]
-\dfrac{40}{7L^{4}\alpha^{2}+20} & 0 
\end{pmatrix}.
\end{eqnarray}
Because the field \eqref{unsteady 2d NS} is composed of a spatially linear, unsteady field which dominates near the origin, and a steady nonlinear field which dominates far away from the origin, we obtain the solution \eqref{matrix T} which consistently leads to the correct identification of the flow type near the origin by setting $L=0$. In contrast, in the limit $L\rightarrow \infty$, the unsteadiness minimization has no effect since the steady, spatially nonlinear flow dominates far from the origin. As we shall see in the next example as well, choosing a sufficiently small domain $\mathcal{D}$ generally improves the accuracy for Eulerian flow analysis since $\bm{T}$ is the dominant part of the unsteadiness of the relative infinitesimal  motion of the flow. 

In the following, we set $\alpha=0.005$ and $L=1$ for which the solution \eqref{Linear equations} reads 
\begin{eqnarray}
    \bm{T}(t_0)=
\begin{pmatrix}
0 & 1.99972 \\
-1.99972 & 0
\end{pmatrix}.
\end{eqnarray}
While the instantaneous streamlines of $\bm{v}$ erroneously predict a vortical flow near the origin, the streamlines of the objective field $\bm{v}_\text{s}=\bm{v}-\bm{T}\bm{x}$ correctly indicate a mixing region near the origin, see Fig. \ref{Figure 1}\textbf{d},\textbf{e}. Choosing larger values for $L$ would, on the contrary, lead to incorrect result. For general $\alpha$ and $L$, the eigenvalues of the auxiliary field $\bm{v}_\text{s}$ are given by 
\begin{eqnarray}
\lambda_{\alpha,\pm}
= \pm\,\frac{\sqrt{-147\,L^{8}\alpha^{4} + 280\,L^{4}\alpha^{2} + 400}}
{7\,L^{4}\alpha^{2} + 20}.
\end{eqnarray}
In contrast, the flow type near the origin is defined by the eigenvalues \eqref{A evs} corresponding to the spatially linear part of \eqref{unsteady 2d NS}, which are given by $\lambda_{\pm} = \pm 1$ in this case. These values exactly correspond to $L=0$ and arbitrary $\alpha$. For sufficiently small $L$ and $\alpha$, the value of $\lambda_{\alpha}$  will note agree with  $\lambda$ exactly, but will still indicate the correct flow type. For this example, both eigenvalue pairs are real, indicating a hyperbolic flow near the origin.
\begin{figure}
    \centering
    \includegraphics[width=1\linewidth]{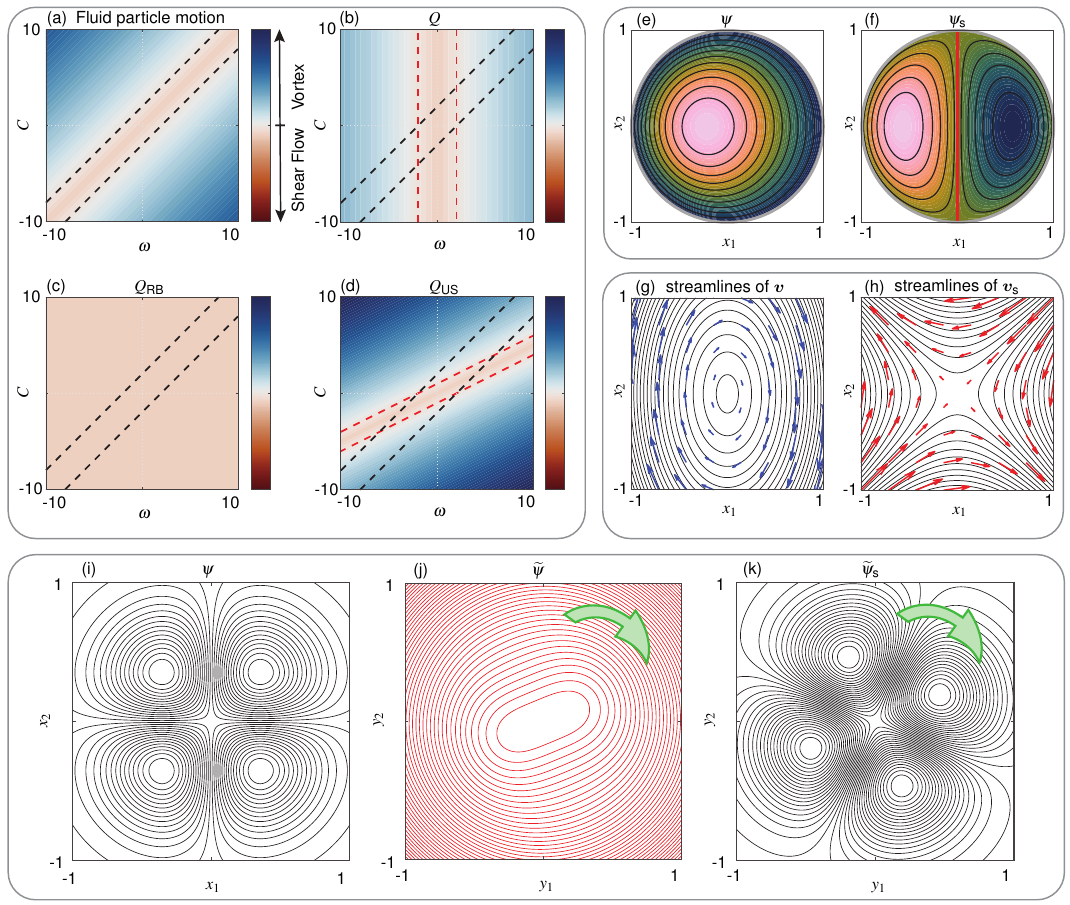}
    \caption{\textbf{Analysis of analytical benchmark examples.} (a), (b), (c), (d): Spatially linear, unsteady Navier--Stokes flow. (e), (f): Separation and reattachment flow. (g), (h): Spatially nonlinear unsteady Navier--Stokes flow. (i), (j), (k): Streamlines corresponding to the four centers flow (i) in the steady, orignal frame, (j) in the rotated, unsteady frame. (k) Streamlines of the objective, auxiliary velocity field $\bm{s}$ of the four centers flow in the rotating, unsteady frame.}
    \label{Figure 5}
\end{figure}

\subsubsection*{Four centers flow}
 The final analytical example has been discussed in \cite{gunther2017generic}, describing a steady flow which is made unsteady through a rotating observer change. In the steady frame, the streamlines indicate the ground truth of the Lagrangian motion defined by the flow, see Fig. \ref{Figure 5}(i), given by four symmetric vortices around the origin. The corresponding streamfunction is given by
 \begin{eqnarray}
     \psi(\bm{x})=x_1\, x_2 \,\mathrm{e}^{-(x_1^{2}+x_2^{2})}
 \end{eqnarray}
In the rotating frame defined by an Euclidean frame change \eqref{Euclidian frame change} with
 \begin{eqnarray}
     \bm{Q}=\begin{pmatrix}
\cos \omega t & \sin  \omega t\\
-\sin  \omega t & \cos  \omega t
\end{pmatrix}
 \end{eqnarray}
 and $\bm{b}=\bm{0}$, the streamlines deceive the observer by showing only a single vortex, which is rotating due to the coordinate change, see Fig. \ref{Figure 5}(j). The streamfunction in the unsteady frame is given by 
 \begin{eqnarray}
     \widetilde{\psi}(\bm{y},t)=\mathrm{e}^{-(y_1^2+y_2^2)}\left[y_1 y_2 \cos(2\omega t)+\frac{1}{2}\left(y_1^2-y_2^2\right)\sin(2\omega t)\right]-\frac{\omega}{2}\left(y_1^2+y_2^2\right).
 \end{eqnarray}
If the unsteadiness minimization, i.e., the computation of $\bm{T}$ described in the Methods section, is performed over the whole real plane, then $\bm{T}$ vanishes identically because the flow is strongly localized near the origin. Motivated by the prior example, we choose instead a vanishingly small, square domain around the origin: $\mathcal{D}=[L,L]^2$, $L\rightarrow 0$. For this choice of domain, the matrix $\bm{T}$ can be computed analytically since the velocity field in the unsteady frame near the origin approximates the spatially polynomial field
\begin{eqnarray}
    \tilde{\bm{v}}(\bm{y},t)\approx \begin{pmatrix}
    y_1\cos(2\theta) + y_2\sin(2\theta) - \omega y_2 \\
  y_1\sin(2\theta) - y_2\cos(2\theta) + \omega y_1
\end{pmatrix}.
\end{eqnarray}
The result is then
\begin{eqnarray}
    \bm{T}=\begin{pmatrix}
        0 & -\omega\\
        \omega & 0
    \end{pmatrix}.
\end{eqnarray}
The streamfunction of the incompressible auxiliary velocity field is given by 
\begin{eqnarray}
\widetilde{\psi}_\text{s}=\widetilde{\psi}(\bm{y},t)+\frac{\omega}{2}(y_1^2+y_2^2).
\end{eqnarray}
The streamlines of $\widetilde{\psi}_\text{s}$ correctly indicate four vortices around the origin, which are rotating due to the coordinate change, and correspond exactly to the ground truth of the steady field up to a change of coordinates. Computing the normed modified unsteady rate $\|\overset{\circ}{\bm{S}}(\bm{x},t;\bm{T}(t))\|$ over the entire real plane yields a scalar field which is maximal at the origin. In this kinematic example of a localized, spatially nonlinear flow, selecting a small domain around the maximum of $\|\overset{\circ}{\bm{S}}(\bm{x},t;\bm{T}(t))\|$ thus leads to the correct result, which further supports the iterative procedure described before. 
 
\subsection*{Computational effort}
\begin{figure}
    \centering
    \includegraphics[width=0.5\linewidth]{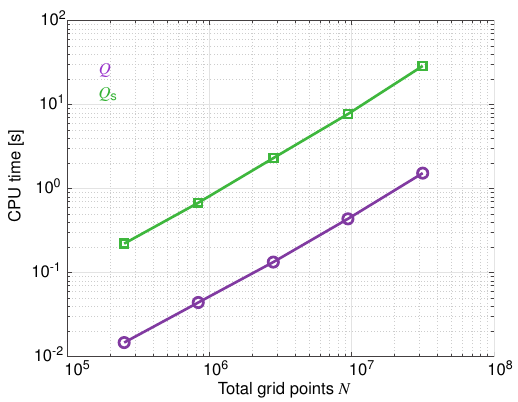}
    \caption{\textbf{Worst-case comparison of computational effort between the $Q$-criterion (purple) and the objective $Q_\text{s}$-criterion (green)} We show the CPU time as a function of the total number of grid points $N$ in a three-dimensional domain. The quantities $Q$ and $Q_\text{s}$ were computed over the domain $[-2,2]^3$ with increasing resolution for the field \eqref{unsteady 2d NS} modified by an additional, constant $z$-velocity component.  The trace identity $Q=-\frac{1}{2}\operatorname{tr}\big[\left(\bm{\nabla}\bm{v}\right)^2\big]$ was used in the computation of $Q$ but not in the computation of $Q_\text{s}$.}
    \label{Figure 7}
\end{figure}

Owing to its Eulerian character, the $Q_{\text{s}}$-criterion is exceptionally efficient and simple to implement, and thus applicable even to challenging, three-dimensional flow data sets for which only a limited number of time steps are available. In particular, no time integration for particle advection is required for this method. Indeed, as is evident from the Methods section, the computational effort to calculate the $Q_\text{s}$-criterion scales like the effort to compute the partial time derivative, with a minor increase due to the need for the computation of $\bm{T}$. However, there is one significant drawback regarding the computational effort of the method presented in this work: Through its definition, the auxiliary velocity field \eqref{objective velocity field} will be compressible whenever $\bm{T}$ is not trace-free. Although $\bm{T}$ is indeed trace-free in all the analytical examples considered before, this is might not hold true in general. Therefore, the identity  $Q=-\frac{1}{2}\operatorname{tr}\big[\left(\bm{\nabla}\bm{v}\right)^2\big]$ which holds for incompressible flows for the $Q$-criterion may not be used generally in the computation of $Q_\text{s}$. To illustrate this worst-case scenario quantitatively, where $Q$ can be computed using the trace identity but $Q_\text{s}$ cannot, we refer to Fig. \ref{Figure 7}: The CPU time required to compute $Q$ and $Q_\text{s}$ is shown against the number of grid points for a three-dimensional Navier--Stokes velocity field which is obtained by adding a constant $z$-velocity component $w=1$ to the two-dimensional field \eqref{unsteady 2d NS}. These calculations were carried out in MATLAB on a Lenovo P16v laptop with a 13th Gen Intel(R) Core(TM) i7-13800H (2.50 GHz) processor without parallelization. The identity $Q=-\frac{1}{2}\operatorname{tr}\big[\left(\bm{\nabla}\bm{v}\right)^2\big]$, although it holds for $Q_\text{s}$ as well in this case, was only used in the computation of $Q$. The matrix $\bm{T}$ was computed once over the three-dimensional domain $[-1,1]^3$ with increasing resolution. The results show a constant offset between the computational effort for the two methods but the same scaling with the number of grid points. We note that one could define an incompressible auxiliary velocity field  from $\bm{T}$ by subtracting only $\operatorname{skew}[\bm{T}]\bm{x}$ from $\bm{v}$ instead of $\bm{T} \bm{x}$. This alternative method, although computationally favorable in this setting, has not been investigated in this work, because it lacks a solid physical interpretation. 

%\subsection*{Data availability}
%Source data are provided with this paper. The data that support the findings of this study are available at \url{https://polybox.ethz.ch/index.php/s/CLxLxoDsWkSuIgq}. 
%\subsection*{Code availability}
%The codes that support the findings of this study are available at \url{https://polybox.ethz.ch/index.php/s/CLxLxoDsWkSuIgq}.

\subsection*{Acknowledgments.} The two data sets analyzed in Section \ref{Example Section} describing a heated flow around a cylinder and the wind wake behind the vessel Tangaroa are both part of the open-source flow database provided by the Computer Graphics Laboratory at ETH Z\"{u}rich. The colormaps used in this work were adopted from \cite{Crameri2020}. The authors would like to thank Bill Kuo, Wei Wang, Cindy Bruyere, Tim Scheitlin, and Don Middleton of the U.S. National Center for Atmospheric Research (NCAR) and the U.S. National Science Foundation (NSF) for providing the Weather Research and Forecasting (WRF) Model simulation data of Hurricane Isabel.\\

\noindent The work of T.P. was supported by the Swiss National Science Foundation under Grant Agreement No. 225619.\\

%\subsection*{Author Contributions}
%\noindent T.P. conceptualized the research. T.P. and F.K. developed the theory. T.P. performed data analysis and validated the method. T.P. and F.K. discussed the results. T.P. wrote the first draft of the article. T.P. and F.K. reviewed and edited it. Both authors approved the final version of the article.

\subsection*{Competing interests}
The authors declare no competing interests.

\bibliography{sn-bibliography}% common bib file
%\input sn-article.bbl

%\label{Appendix_Exp}
%\label{Appendix_AcField}
%\label{Appendix_Fluid}
%\label{Appendix_BifDiag}
%\label{Appendix_Zeeman}
%\label{Appendix_Unbiased}
%\label{Appendix_Sync}

~\\

\clearpage

~\\
~\\
\begin{center}
  \Large Objective detection of coherent vortices from\\ {instantaneous flow data}\\[1cm]
 \large  Tiemo Pedergnana,$^{1*}$ and Florian Kogelbauer$^{2}$ \\[.5cm]
  { ${}^1$Department of Mechanical Engineering,\\ Massachusetts Institute of Technology, MA, USA\\
  ${}^2$Department of Mathematics, ETH Zürich, Zürich, Switzerland}\\[.5cm]
  ${}^*$Corresponding author. E-mail: {tiemop@mit.edu};
\\[1cm]
\end{center}

\baselineskip24pt


\maketitle 
\label{suppmat}
\noindent\textbf{This PDF file includes:}
\begin{enumerate}
         \item[] \textbf{{Supplementary Notes} \label{supptext}}
   \begin{enumerate}
     \item[]  Matrix vectorization
     \item[]  First variation of the action
      \item[]  Transformation property of $\bm{T}$
     \item[]  Spatially linear systems
       \end{enumerate}
     \end{enumerate}
     
\thispagestyle{empty}

\setcounter{equation}{0}
\setcounter{figure}{0}
\setcounter{table}{0}
\setcounter{page}{1}
\renewcommand{\theequation}{S\arabic{equation}}
\renewcommand{\thefigure}{S\arabic{figure}}

\clearpage

\section*{{Supplementary Notes}}
\label{Supplementary Notes}
\subsection*{Matrix vectorization}
This work makes use of the \textit{vectorization} of a  matrix $\bm{A}=[A_{ij}]\in \mathbb{R}^{m\times n}$, denoted as $\operatorname{vec}(\bm{A})$,
which is the $mn\times 1$ column vector obtained by stacking the columns of $\bm{A}$:
\begin{eqnarray}
\bm{A} = 
\begin{pmatrix} 
A_{11} & A_{12} & \dots & A_{1n} \\ 
A_{21} & A_{22} & \dots & A_{2n} \\ 
\vdots & \vdots & \ddots & \vdots \\ 
A_{m1} & A_{m2} & \dots & A_{mn} 
\end{pmatrix}
\quad \Rightarrow \quad
\operatorname{vec}(\bm{A}) =
\begin{pmatrix} 
A_{11} \\ A_{21} \\ \vdots \\ A_{m1} \\ A_{12} \\ \vdots \\ A_{mn} 
\end{pmatrix}.
\end{eqnarray} 
We denote the standard inner product on the space of matrices as $\langle \bm{A},\bm{B} \rangle = \text{tr} (\bm{A}^T \bm{B})$ with corresponding norm $\|\bm{A}\|=\sqrt{\langle \bm{A},\bm{A}\rangle}$ and recall the identity  $\mathrm{tr} \left(\bm{A}^{T}\bm{B}\right) =\operatorname{vec} \left[\bm{A}\right]^{T}\,\operatorname{vec} \left[\bm{B}\right]$,  see \cite{magnus2007matrix}.

\subsection*{First variation of the action \label{first variation}}
We compute the first variation of the action defined in \eqref{action} as
\begin{eqnarray}
     \frac{\delta \mathcal{S}}{\delta \bm{T}}\hat{\bm{T}} 
      & =& \frac{d}{d\varepsilon}\Big\vert_{\varepsilon = 0} \mathcal{S}[\bm{T} + \varepsilon \hat{\bm{T}}]\\
      & =&  \frac{d}{d\varepsilon}\Big\vert_{\varepsilon = 0} \frac{1}{2}\int_{t_0}^{t_1}\fint_{\mathcal{D}} \|\partial_t \bm{S}-(\bm{T}+\varepsilon \hat{\bm{T}})\bm{S}-\bm{S}(\bm{T}+\varepsilon \hat{\bm{T}})^T\|^2\, dV dt\\
      & =&   \int_{t_0}^{t_1}\fint_{\mathcal{D}} \mathrm{tr}\left[(\hat{\bm{T}}\bm{S}+\bm{S}\hat{\bm{T}}^T)^T (\bm{T}\bm{S}+\bm{S}\bm{T}^T-\partial_t \bm{S})\right]\, dV dt.\\
&=&\int_{t_0}^{t_1}\fint_{\mathcal{D}}\operatorname{vec}[\hat{\bm{T}}]^{T}\,\bm{M}^{T}\left(\bm{M}\,\operatorname{vec}[\bm{T}]-\operatorname{vec}[\partial_t \bm{S}]\right)\, dV\, dt,
\end{eqnarray}
where $\bm{M} = \bm{S}^{T} \otimes \bm{I}
+ (\bm{I} \otimes \bm{S})\,\bm{K}$ and $\bm{K}$ is the commutation matrix satisfying $\operatorname{vec}[\bm{A}^T]=\bm{K}\,\operatorname{vec}[\bm{A}]$. Using $\bm{S}^T=\bm{S}$ and setting the first variation to zero yields
\begin{eqnarray} \label{Solution of minization problem}
     \frac{\delta \mathcal{S}}{\delta \bm{T}}= 
     \overline{\bm{M}^{T}\left(\bm{M}\,\operatorname{vec}[\bm{T}]-\operatorname{vec}[\partial_t \bm{S}]\right)}=\bm{0}.
\end{eqnarray}
Assuming the spatial average $\overline{\bm{M}^{T}
\bm{M}}$ to be non-singular, the solution of this equation is given by \eqref{Linear equations}. The domain-dependence and the (non-)singularity of $\overline{\bm{M}^{T}
\bm{M}}$ is examined in detail in the Discussion section of the main text.

\subsection*{Transformation property of $\bm{T}$} \label{Appendix A}
In this appendix, we derive the transformation property of the matrix $\bm{T}$, which appears in the modified rate of the rate-of-strain tensor $\bm{S}$ defined in \eqref{Objective rate}, under a general Euclidean frame change \eqref{Euclidian frame change}. This derivation makes use of the transformation property of $\bm{S}$, which is itself an objective tensor, see \cite{TruesdellNoll2004}, and its time derivative:
\begin{eqnarray}
     \widetilde{\bm{S}} &=& \bm{Q}^{T}\bm{S}\bm{Q} 
\quad \Rightarrow \quad \widetilde{\bm{S}}^{T} = \bm{Q}^{T}\bm{S}^{T}\bm{Q},\\
\widetilde{\partial_t \bm{S}}&=&\bm{Q}^T\partial_t \bm{S}\bm{Q}+\dot{\bm{Q}}^T\bm{S}\bm{Q}+\bm{Q}^T\bm{S}\dot{\bm{Q}}  \notag \\
&=&\bm{Q}^T\left(\partial_t \bm{S}-\dot{\bm{Q}}\bm{Q}^T\bm{S}-\bm{S}\bm{Q}\dot{\bm{Q}}^T\right)\bm{Q},
\end{eqnarray}
where {we used $\widetilde{\bm{v}}_{\text{d}}=\bm{Q}^T \bm{v}_{\text{d}}$} and the property that {$\dot{\bm{Q}} = -\bm{Q}\dot{\bm{Q}}^T\bm{Q}$} for the time-derivative of rotation matrices. We also recall the following three identities for the Kronecker product and the vectorization operator. 
For square matrices 
$\bm{A}, \bm{B}, \bm{C}, \bm{M}, \bm{N}, \bm{P}, \bm{Q}, \bm{I} \in \mathbb{R}^{n\times n}$, 
we have
\begin{eqnarray}
    (\bm{M}\bm{A}\bm{P}) \otimes (\bm{N}\bm{B}\bm{Q})
    &=& (\bm{M}\!\otimes\!\bm{N})\,(\bm{A}\!\otimes\!\bm{B})\,(\bm{P}\!\otimes\!\bm{Q}),\\
    \operatorname{vec}(\bm{A}\bm{B}\bm{C})
    &=& (\bm{C}^{T}\!\otimes\!\bm{A})\,\operatorname{vec}(\bm{B})
    = (\bm{I}\!\otimes\!\bm{A})(\bm{C}^{T}\!\otimes\!\bm{I})\,\operatorname{vec}(\bm{B}),\\
     (\bm{A} \otimes \bm{B})^T &=&  \bm{A}^T \otimes \bm{B}^T.
\end{eqnarray}
The transformation property of an objective matrix $\bm{A}$ under a frame change \eqref{Euclidian frame change} is 
\begin{eqnarray}
\operatorname{vec}[\widetilde{\bm{A}}] &=& \operatorname{vec}[\bm{Q}^T \bm{A} \bm{Q}]= (\bm{Q}^T \!\otimes\! \bm{Q}^T)\, \operatorname{vec}[\bm{A}].
\end{eqnarray}

\subsubsection*{Transformation property of ${\bm{M}}$}
Using the fact that the commutation matrix $\bm{K}$ satisfies $\bm{K} \, (\bm{A} \otimes \bm{B}) 
= (\bm{B} \otimes \bm{A}) \, \bm{K}$, we calculate the transformation property of ${\widetilde{\bm{M}}}$ under an Euclidean frame change: 
\begin{eqnarray}
    {\widetilde{\bm{M}}} &=& {\widetilde{\bm{S}}^{T}\!\otimes\!\bm{I} + (\bm{I}\!\otimes\!\widetilde{\bm{S}})\bm{K}} \notag  \\
&=& {(\bm{Q}^{T}\bm{S}^{T}\bm{Q})\!\otimes\!\bm{I} 
   + (\bm{I}\!\otimes\!\bm{Q}^{T}\bm{S}\bm{Q})\bm{K}}  \notag \\
&=& {(\bm{Q}^{T}\!\otimes\!\bm{I})(\bm{S}^{T}\!\otimes\!\bm{I})(\bm{Q}\!\otimes\!\bm{I})
   + (\bm{I}\!\otimes\!\bm{Q}^{T})(\bm{I}\!\otimes\!\bm{S})(\bm{I}\!\otimes\!\bm{Q})\bm{K}} \notag \\
   &=& {(\bm{Q}^{T}\!\otimes\!\bm{Q}^T)(\bm{S}^{T}\!\otimes\!\bm{I})(\bm{Q}\!\otimes\!\bm{Q})
   + (\bm{Q}^T\!\otimes\!\bm{Q}^{T})(\bm{I}\!\otimes\!\bm{S})(\bm{Q}\!\otimes\!\bm{Q})\bm{K}} \notag \\
    &=& {(\bm{Q}^{T}\!\otimes\!\bm{Q}^T)(\bm{S}^{T}\!\otimes\!\bm{I})(\bm{Q}\!\otimes\!\bm{Q})
   + (\bm{Q}^T\!\otimes\!\bm{Q}^{T})(\bm{I}\!\otimes\!\bm{S})\bm{K}(\bm{Q}\!\otimes\!\bm{Q})} \notag \\
&=&(\bm{Q}^T\!\otimes\!\bm{Q}^T){{\bm{S}}^{T}\!\otimes\!\bm{I} + (\bm{I}\!\otimes\!{\bm{S}})\bm{K}}(\bm{Q}\!\otimes\!\bm{Q}) \notag \\
 &=& (\bm{Q}^{T}\!\otimes\!\bm{Q}^{T})\,{\bm{M}}\,(\bm{Q}\!\otimes\!\bm{Q}). \label{intapp}
\end{eqnarray}
and, consequently, 
\begin{eqnarray}{\widetilde{\bm{M}}^{T}\widetilde{\bm{M}}} &=& (\bm{Q}^{T}\!\otimes\!\bm{Q}^{T})\,{\bm{M}^{T}\bm{M}}\,(\bm{Q}\!\otimes\!\bm{Q}).
\end{eqnarray}

\subsubsection*{Transformation property of $\overline{\bm{M}^{T}\operatorname{vec}[\partial_t \bm{S}]}$}
The transformation property \eqref{intapp} implies that we can write the transformed spatial average as
\begin{eqnarray}
\overline{\widetilde{\bm{M}}^{T} \, \operatorname{vec}[\widetilde{\partial_t \bm{S}}]}
&=& \overline{ (\bm{Q} \otimes \bm{Q})^{T} \, \bm{M}^T \, (\bm{Q}^T \otimes \bm{Q}^T)^T 
\, \operatorname{vec} \Big[ \bm{Q}^T \partial_t \bm{S} \bm{Q} + \dot{\bm{Q}}^T \bm{S} \bm{Q} + \bm{Q}^T \bm{S} \dot{\bm{Q}} \Big] }\notag\\
&=& \overline{ (\bm{Q}^T \!\otimes\! \bm{Q}^T) \, \bm{M}^T \, (\bm{Q} \!\otimes\! \bm{Q}) 
\, \operatorname{vec} \Big[ \bm{Q}^T \partial_t \bm{S} \bm{Q} + \dot{\bm{Q}}^T \bm{S} \bm{Q} + \bm{Q}^T \bm{S} \dot{\bm{Q}} \Big] }. \notag \\
&=& \overline{ (\bm{Q}^T \!\otimes\! \bm{Q}^T) \, \bm{M}^T \, (\bm{Q} \!\otimes\! \bm{Q}) 
\, \operatorname{vec} \Big[ \bm{Q}^T\left(\partial_t \bm{S}-\dot{\bm{Q}}\bm{Q}^T\bm{S}-\bm{S}\bm{Q}\dot{\bm{Q}}^T\right)\bm{Q}\Big]\notag  }.\label{intermediate}
\end{eqnarray}

Treating each term in the square bracket separately, we obtain 
\begin{eqnarray}
(\bm{Q} \!\otimes\! \bm{Q}) \, \operatorname{vec}[\bm{Q}^T \partial_t \bm{S} \bm{Q}]
&=& (\bm{Q}\bm{Q}^T \!\otimes\! \bm{Q}\bm{Q}^T) \, \operatorname{vec}[\partial_t \bm{S}]
= \operatorname{vec}[\partial_t \bm{S}], \notag\\
(\bm{Q} \!\otimes\! \bm{Q}) \, \operatorname{vec}[\dot{\bm{Q}}^T \bm{S} \bm{Q}] 
&=& (\bm{I} \!\otimes\! \bm{Q}\dot{\bm{Q}}^T) \, \operatorname{vec}[\bm{S}] = \operatorname{vec}[\bm{Q}\dot{\bm{Q}}^T\bm{S}], \notag\\
(\bm{Q} \!\otimes\! \bm{Q}) \, \operatorname{vec}[\bm{Q}^T \bm{S} \dot{\bm{Q}}]
&=& (\bm{Q}\dot{\bm{Q}}^T \!\otimes\! \bm{I}) \, \operatorname{vec}[\bm{S}] = \operatorname{vec}[\bm{S}\dot{\bm{Q}}\bm{Q}^T  ].
\end{eqnarray}
Combining the above results gives
\begin{eqnarray}
(\bm{Q} \!\otimes\! \bm{Q}) \, \operatorname{vec}[\bm{Q}^T \partial_t \bm{S} \bm{Q} + \dot{\bm{Q}}^T \bm{S} \bm{Q} + \bm{Q}^T \bm{S} \dot{\bm{Q}}] 
&=& \operatorname{vec}[\partial_t \bm{S} + {\bm{Q}} \dot{\bm{Q}}^T \bm{S} + \bm{S} \dot{\bm{Q}} {\bm{Q}}^T]. \notag \\
\end{eqnarray}
Using $\dot{\bm{Q}}\bm{Q}^T = -\bm{Q}\dot{\bm{Q}}^T$, we obtain 
\begin{eqnarray}
\overline{\widetilde{\bm{M}}^{T} \, \operatorname{vec}[\widetilde{\partial_t \bm{S}}]} 
&=&(\bm{Q}^T \otimes \bm{Q}^T) \, \overline{  \bm{M}^T \, \operatorname{vec}\big[ \partial_t \bm{S} +{\bm{Q}} \dot{\bm{Q}}^T \bm{S} - \bm{S} \bm{Q}\dot{\bm{Q}}^T \big] }.
\end{eqnarray}

\subsubsection*{Transformation property of $\operatorname{vec}[\bm{T}]$}
By combining the above results, we obtain
\begin{eqnarray}
\operatorname{vec}[\widetilde{\bm{T}}] 
&=& \overline{\widetilde{\bm{M}}^T \widetilde{\bm{M}}}^{-1} \, 
\overline{\widetilde{\bm{M}}^T \operatorname{vec}[\widetilde{\partial_t \bm{S}}]}\notag\\
&=& \Big[ (\bm{Q}^T \otimes \bm{Q}^T) \, \overline{\bm{M}^T \bm{M}} \, (\bm{Q} \otimes \bm{Q}) \Big]^{-1} 
(\bm{Q}^T \otimes \bm{Q}^T) \, 
\overline{ \bm{M}^T \, \operatorname{vec}[\partial_t \bm{S} +{\bm{Q}} \dot{\bm{Q}}^T \bm{S} - \bm{S} \bm{Q}\dot{\bm{Q}}^T] } \notag \\
&=& (\bm{Q}^T \otimes \bm{Q}^T) \, \overline{\bm{M}^T \bm{M}}^{-1} \, 
\overline{ \bm{M}^T \, \operatorname{vec}[\partial_t \bm{S} +{\bm{Q}} \dot{\bm{Q}}^T \bm{S} - \bm{S} \bm{Q}\dot{\bm{Q}}^T] } \notag \\
&=& (\bm{Q}^T \otimes \bm{Q}^T) \, \overline{\bm{M}^T \bm{M}}^{-1} \, 
\overline{ \bm{M}^T \, \operatorname{vec}[\partial_t \bm{S}]} \notag \\
&& - (\bm{Q}^T \otimes \bm{Q}^T) \,  \overline{\bm{M}^T \bm{M}}^{-1} \, 
\overline{ \bm{M}^T \, \operatorname{vec}[\bm{S} \bm{Q}\dot{\bm{Q}}^T-{\bm{Q}} \dot{\bm{Q}}^T \bm{S}  ] }. \label{substitute here}
\end{eqnarray}
Using the fact that 
\begin{eqnarray}
\bm{M}\,\operatorname{vec}[\dot{\bm{Q}} {\bm{Q}}^T] 
&=& (\bm{S}^T \otimes \bm{I} + (\bm{I} \otimes \bm{S}) \, \bm{K}) \, \operatorname{vec}[\dot{\bm{Q}} {\bm{Q}}^T] \notag\\
&=& (\bm{S}^T \otimes \bm{I}) \, \operatorname{vec}[\dot{\bm{Q}} {\bm{Q}}^T] \;+\; (\bm{I} \otimes \bm{S}) \, \bm{K} \, \operatorname{vec}[\dot{\bm{Q}} {\bm{Q}}^T] \notag\\
&=& \operatorname{vec}[(\dot{\bm{Q}} {\bm{Q}}^T) \bm{S}] \;+\; \operatorname{vec}[\bm{S} (\dot{\bm{Q}} {\bm{Q}}^T)^T] \notag\\
&=& \operatorname{vec}[\dot{\bm{Q}}\bm{Q}^T  \bm{S} \;+\; \bm{S}  \bm{Q} \dot{\bm{Q}}^T] \notag\\
&=&\operatorname{vec}[\bm{S}  \bm{Q} \dot{\bm{Q}}^T \;-\; \bm{Q}\dot{\bm{Q}}^T  \bm{S}] ,
\end{eqnarray}
we can further simplify:
\begin{eqnarray}
(\bm{Q}^T \!\otimes\! \bm{Q}^T) \, \overline{\bm{M}^T \bm{M}}^{-1} \, 
\overline{ \bm{M}^T \, \operatorname{vec}[\bm{S}  \bm{Q} \dot{\bm{Q}}^T \;-\; \bm{Q}\dot{\bm{Q}}^T  \bm{S} ] } &=& (\bm{Q}^T \!\otimes\! \bm{Q}^T) \, \overline{\bm{M}^T \bm{M}}^{-1} \, 
\overline{ \bm{M}^T \bm{M}} \, \operatorname{vec}[\dot{\bm{Q}} \bm{Q}^T]  \notag \\
&=& (\bm{Q}^T \!\otimes\! \bm{Q}^T) \, 
\operatorname{vec}[\dot{\bm{Q}} \bm{Q}^T] \notag \\
&=& \operatorname{vec}[ \bm{Q}^T (\dot{\bm{Q}} \bm{Q}^T) \bm{Q} ] \notag \\
&=& \operatorname{vec}[ \bm{Q}^T\dot{\bm{Q}} ],
\end{eqnarray}
By substituting this result back into \eqref{substitute here}, we have shown that the matrix $\bm{T}$ indeed transforms like a spin tensor under a general Euclidean frame change. 

\subsection*{Spatially linear fields}
For spatially linear fields, $\bm{M}$ and $\bm{S}$ are independent of $\bm{x}$ and $\overline{\bm{M}^{T}
\bm{M}}=\bm{M}^{T}
\bm{M}$ is always singular. In this case, we need to use the Moore--Penrose pseudoinverse in the least-squares solution formula \eqref{Linear equations pseudo}. In this case,
\begin{eqnarray} \label{MP Pseudoinverse appendix}
    \operatorname{vec}[\bm{T}_\text{lsq}]
&=&\overline{\bm{M}^{T}
\bm{M}}^{+}\,\overline{\bm{M}^{T}\operatorname{vec}[\partial_t \bm{S}]}\\
&=&\left(\bm{M}^{T}
\bm{M}\right)^{+}\,\bm{M}^{T}\operatorname{vec}[\partial_t \bm{S}]\\
&=&\bm{M}^+\left(\bm{M}^{T}\right)^+
\,\bm{M}^{T}\operatorname{vec}[\partial_t \bm{S}]\\
&=&\bm{M}^+\operatorname{vec}[\partial_t \bm{S}].
\end{eqnarray}
where we used $(\bm{A}^T\bm{A})^+=\bm{A}^+(\bm{A}^T)^+$, $(\bm{A}^T)^+=(\bm{A}^+)^T$ and the symmetry of $\bm{M} \bm{M}^+$, see \cite{magnus2007matrix}, which lead to
\begin{eqnarray}
   \bm{M}^+ (\bm{M}^T)^+ \bm{M}^T = \bm{M}^+ (\bm{M}^+)^T \bm{M}^T = \bm{M}^+ (\bm{M} \bm{M}^+)^T = \bm{M}^+ (\bm{M} \bm{M}^+) = \bm{M}^+.\notag \\
\end{eqnarray}
Using the transformation property of $\bm{M}$ derived in \eqref{intapp} and $(\bm{Q}^T\overline{\bm{M}^T\bm{M}}\bm{Q})^{+}=\bm{Q}^T \overline{\bm{M}^T\bm{M}}^{+} \bm{Q}$ for orthogonal matrices $\bm{Q}$, we then obtain
\begin{eqnarray} \label{Transformation MPP}
    \widetilde{\operatorname{vec}[\bm{T}]}
&=&\widetilde{\bm{M}^+}\widetilde{\operatorname{vec}[\partial_t \bm{S}]}\notag\\
&=&\left[(\bm{Q}^T \otimes \bm{Q}^T) \, 
 \bm{M} \, (\bm{Q} \otimes \bm{Q})\right]^+(\bm{Q}^T \otimes \bm{Q}^T)\operatorname{vec}[\partial_t \bm{S} - \dot{\bm{Q}}^T \bm{Q} \bm{S} - \bm{S} \dot{\bm{Q}} \bm{Q}^T] \notag\\
&=&(\bm{Q}^T \otimes \bm{Q}^T) \, 
{ \bm{M}^+ \,\operatorname{vec}[\partial_t \bm{S} - \dot{\bm{Q}}^T \bm{Q} \bm{S} - \bm{S} \dot{\bm{Q}} \bm{Q}^T] }\notag\\
&=&(\bm{Q}^T \otimes \bm{Q}^T) \, 
 \bm{M}^+ \, \operatorname{vec}[\partial_t \bm{S}]-(\bm{Q}^T \!\otimes\! \bm{Q}^T)\bm{M}^+\bm{M}\,\operatorname{vec}[\dot{\bm{Q}} \bm{Q}^T].\notag
\end{eqnarray}
Therefore, the least-squares solution \eqref{Linear equations pseudo} transforms objectively if 
\begin{eqnarray}
    (\bm{Q}^T \!\otimes\! \bm{Q}^T)\bm{M}^+\bm{M}\,\operatorname{vec}[\dot{\bm{Q}} \bm{Q}^T]=\operatorname{vec}[ \bm{Q}^T\dot{\bm{Q}} ].
\end{eqnarray}




\end{document}